\documentclass[10pt,twocloumn]{IEEEtran}
\usepackage{bbding}
\usepackage{bbm}
\usepackage{amsfonts}
\usepackage{array}
\usepackage{amsmath}
\usepackage{graphicx}
\usepackage[usenames,dvipsnames]{color}
\usepackage{multirow,footnote,array,booktabs}
\usepackage{fancybox}
\usepackage[dvipsnames,svgnames,x11names]{xcolor}
\usepackage{graphicx}
\usepackage{cite}
\usepackage{url}
\usepackage{amsthm}
\usepackage{epsfig, psfrag, amsmath, amssymb, amsfonts, latexsym, eucal, balance, hyperref, indentfirst,bookmark, amsthm, dsfont}
\usepackage{graphicx}
\usepackage{subfigure}
\usepackage{verbatim}
\usepackage{soul}
\soulregister\cite7
\usepackage{ulem}

\newtheorem{theorem}{Theorem}
\newtheorem{definition}{Definition}
\newtheorem{lemma}{Lemma}

\newtheorem{approximation}{Approximation}
\newtheorem{rem}{Remark}

\graphicspath{{figs/}}

\ifCLASSINFOpdf
\else
\fi

\begin{document}
\title{Effect of Spatial and Temporal Traffic Statistics on the Performance of Wireless Networks}
\author{Gang Wang, \normalem\emph{Student Member, IEEE}, Yi Zhong, \normalem\emph{Member, IEEE}, Rongpeng Li, \emph{Member, IEEE},\\ Xiaohu Ge, \emph{Senior Member, IEEE}, Tony Q.S. Quek, \emph{Fellow, IEEE}, and Guoqiang Mao, \emph{Fellow, IEEE}
\thanks{
Gang Wang, Yi Zhong and Xiaohu Ge are with
School of Electronic Information and Communications, Huazhong University
of Science and Technology, Wuhan, China.
Rongpeng Li is with Zhejiang University, Hangzhou, China.
Tony Q.S. Quek is with Information Systems Technology and Design, Singapore University of Technology and Design, Singapore.
Guoqiang Mao is with School of Telecommunications Engineering, Xidian University, Xian, China.

This research was supported by the National Natural Science Foundation of China (NSFC)
grants No. 61701183 and No. 61701439, and in part by the Fundamental Research Funds for the Central Universities through grant 2018KFYYXJJ139 and the Cheng Tsang Man Chair Professorship. Part of this work was presented at the 2018 IEEE Global Communications Conference (GLOBECOM) \cite{gangGlobecom}.

The corresponding author is Yi Zhong (e-mail: yzhong@hust.edu.cn).
}
}


\maketitle
\pagestyle{empty}  
\thispagestyle{empty} 

\begin{abstract}
The traffic in wireless networks has become diverse and fluctuating both spatially and temporally due to
the emergence of new wireless applications and the complexity of scenarios. The purpose of this paper is to quantitatively analyze the impact of the wireless traffic, which fluctuates both spatially and temporally, on the performance of the wireless networks.  Specially, we propose to combine the tools from stochastic geometry and queueing theory to model the spatial and temporal fluctuation of traffic, which to our best knowledge has seldom been evaluated analytically.  We derive the spatial and temporal statistics, the total arrival rate, the stability of queues and the delay of users by considering two different spatial properties of traffic, i.e., the uniformly and non-uniformly distributed cases.  The numerical results indicate that although the fluctuation of traffic (reflected by the variance of total arrival rate) when the users are clustered is much fiercer than that when the users are uniformly distributed, the unstable probability is smaller. Our work provides a useful reference for the design of wireless networks when the complex spatio-temporal fluctuation of the traffic is considered.
\end{abstract}



\section{Introduction}
\subsection{Motivations}
The emergence of various smart devices and wireless applications, such as real-time wireless gaming, smart grid, free-viewpoint video, advanced manufacturing and Tactile Internet \cite{7403840, 6755599}, has led to diversified traffic and quality-of-service (QoS) requirements.
For example, the voice traffic in the wireless networks is typically delay-sensitive and symmetric in uplink and downlink, while the data and video traffic are generally loss-sensitive and asymmetric in uplink and downlink, which are IP-based and can tolerate certain delay \cite{mobile2011global}.
A meaningful and practically relevant problem is to meet the QoS requirements of diversified applications, which is also one of the most significant goals for the 5G wireless networks.


With the continuous evolution of wireless networks, the tremendous traffic and its dynamic variations have become more and more significant in affecting the performance of wireless networks.
The pattern of the traffic in a wireless network determines whether a resource block of a base station (BS) is occupied or not, which then shapes the interference pattern in the wireless network.
The interference pattern, in turn, affects the performance of the transceiver links in the wireless network, which directly determines the service process of the traffic.
Therefore, the traffic and the service provided by the wireless networks are highly coupled with each other. Modeling and analysis of the traffic are essential to design and configure the wireless networks so as to match the network service with the traffic \cite{8436053}.


The spatial distribution and the temporal variation of wireless traffic are much more complicated than before.
For example, in a cellular network, the wireless traffic during a holiday or a weekend is generally lighter than that during a weekday, and the traffic during the midnight is generally lighter than that during the day time.
Meanwhile, the spatial distributions of the wireless traffic are also very different between different regions.
For example, the traffic in the business regions could be much heavier
during the day time than that during the midnight, and it is reversed in the residential region.
The integrated analysis of the traffic with spatial and temporal fluctuation requires to appropriately model both the spatial distribution and the temporal variation of the traffic in the wireless networks. Although the effect of the wireless traffic has been studied extensively, most of them only consider one aspect of the traffic \cite{6757900,Wang2757518,7890496}.
The works only modeling the spatial distribution usually use the tools from the stochastic geometry and model the spatial distribution of users by either the uniform (such as the Poisson point process (PPP)) or the non-uniform point processes. Other works only modeling the temporal variations of traffic usually use the queueing theory to model the arrival process of the packets as stochastic processes. To model both aspects of the traffic, which is necessary for analyzing the wireless networks, requires the combination of stochastic geometry and queueing theory, which brings in more complexities and difficulties in the modeling and evaluation.

\subsection{Related Works}
Related works are summarized as follows. The authors in \cite{6757900,Wang2757518} discussed the patterns and the spatio-temporal characteristics of wireless traffic in the practical cellular networks. In \cite{6757900}, the authors presented the analysis of traffic measurements collected from commercial cellular networks in China and proposed a spatial traffic model which generated large-scale spatial traffic variations by a sum of sinusoids. In \cite{Wang2757518}, the authors quantitatively characterized the spatio-temporal distribution of mobile traffic and presented a detailed visualized analysis, and the work \cite{6056689} revealed that the traffic is typically unbalanced, changing not only in the time domain but also in the spatial domain. In \cite{7762185}, the authors extracted and modeled the traffic patterns of large scale towers deployed in a metropolitan city.
In \cite{Cici2746292}, the authors showed some cell phone activity patterns based on the cell phone data which consists of telecommunications activity records in the city of Milan from Telecom Italia Mobile, and the patterns demonstrated that the mobile traffic of urban ecology were clustered in both the time domain and the spatial domain. The above works reveal that both the spatial distribution and the temporal variation of traffic are irregular, and there is a clear need for new analytical methods to explore the properties of irregular distributions and variations of the up-to-date wireless traffic.

Theoretical analysis of the spatial properties of the wireless traffic is generally based on the stochastic geometry.
In \cite{6516167}, the probability density function (PDF) of the number of users in each Voronoi cell was derived by modeling the locations of users as a homogeneous PPP. The works \cite{6497002,6287527} extended such PDFs to the case of multi-tier heterogeneous networks. In \cite{6497002}, the authors analyzed the effect of the offloading traffic in heterogeneous networks on the system performance and the authors in \cite{6287527} developed a framework to characterize the signal-to-noise-plus-interference ratio (SINR) in a heterogeneous cellular network.
In\cite{8168355}, the authors considered a single-tier uplink Poisson cellular network and analyzed the meta distribution of the SIR for both the cellular network uplink and downlink with fractional power control. In \cite{5226957,haenggi2012stochastic}, the authors summarized the applications of the point processes in wireless networks, where the average behavior over many spatial realizations of a network can be evaluated appropriately. In \cite{8267238}, the authors proposed a model to capture the coupling between users and small cell base stations by using tools from stochastic geometry.
In \cite{6042301}, the authors compared the traditional square grid model with the PPP model in terms of coverage, and they discussed the average achievable rate in the general case.
As for the temporal arrival of packets, the analysis was generally based on the queueing theory.
In\cite{8405771}, the authors considered a shared access network with one primary source-destination pair and many secondary communication pairs
and they discussed secondary throughput optimization with primary delay constraints by using tools from stochastic geometry and queueing theory.
In\cite{8281002}, the authors investigated the effect of bursty traffic and analyzed the delay and throughput in a wireless caching system by using queueing theory.
The queueing delay was analyzed with a random access network in \cite{8303680}.
In \cite{21216,tsybakov1979ergodicity,szpankowski1994stability,79909,luo1999stability}, a discrete-time slotted ALOHA system with multiple terminals was described, where each terminal had an infinite buffer to store the packets.


The works modeling both the spatial and temporal aspects of the wireless traffic include \cite{7417573, 7417237, yzhong2020jsac}, where the authors simultaneously modeled the spatio-temporal arrival of traffic and considered the traffic generated at random spatial regions. In \cite{7886285}, the authors discussed three kinds of scheduling policies, i.e., the random scheduling, the first-input-first-output (FIFO) scheduling and the round-robin scheduling, and compared the delay performance under different scheduling policies.
In\cite{7917340}, the authors developed a traffic-aware spatio-temporal mathematical model for Internet of Things (IoT) devices supported by cellular uplink connectivity and discussed the stability for three transmission strategies.
In\cite{8408843}, the authors analyzed the random access channel in celluar-based massive IoT networks based on a traffic-aware spatio-temporal model, where the spatial topology was modeled based on stochastic geometry and the queue evolution was analyzed based on probability theory.
In\cite{7842367}, the authors proposed a user-centric mobility management to cope with user spatial movement and temporally correlated wireless channels in ultra-dense cellular.
In \cite{8335767}, the authors evaluated the tradeoff between delay and physical layer security.
In \cite{blaszczyszyn}, the authors combined stochastic geometry and queueing theory to describe the network with spatial irregularity and temporal evolution.
However, the non-uniform property of network and the various temporal arrival processes of traffic are not considered.

\subsection{Contributions}
In this paper, we establish a tractable model to characterize both the spatial distribution and the temporal variation of traffic in wireless networks. The spatial distribution of traffic could be described by the locations of users. Ignoring the mobility of users, the temporal variation of traffic is described by a random arrival process of packets for each user and the dynamic serving process of arrived packets. We consider two different packet arrival rate distributions, i.e., uniform distribution and exponential distribution. Considering both the uniformly distributed traffic modeled by a PPP and the non-uniformly distributed traffic modeled by a Poisson cluster process (PCP), we explore the relationship between traffic and delay to gain insight. The main contributions of this paper are summarized as follows.

\begin{itemize}
\item
Analytical framework based on combining stochastic geometry and queueing theory is proposed to qualitatively evaluate the spatial and temporal fluctuation of the wireless traffic.

\item
The spatial and temporal statistics of traffic, the stability of queues and the delay are derived for uniformly and non-uniformly distributed traffic. The numerical analyses based on the theoretical results are investigated to gain insights.

\item
The effect of various parameters of the traffic is discussed in terms of meeting the delay and stability requirements.
Our results reveal that although the fluctuation of traffic (reflected by the variance of total arrival rate) when the users are clustered is much fiercer than that when the users are uniformly distributed, the unstable probability is smaller.
\end{itemize}

\begin{table}
    \centering
    \caption{Notations Used in this paper}
    \label{notation_definition}
    \begin{tabular}{|c|c|}
        \hline
        \textbf{Notation} & \textbf{Definition}\\
        \hline
        {$\Phi_b$}& PPP of BSs with density $\lambda_b$ \\
        \hline
        {$\Phi_u$}& Point process of users with density $\lambda_u$ \\
        \hline
        {$P_b$}& Transmit power \\
        \hline
        {$\Phi _p$}& Parent point process with density $\lambda_p$ \\
        \hline
        {$\Phi _x$}& Daughter point process ($x \in {\Phi _p}$) with density $\lambda_c$ \\
        \hline
        {$r_c$}& Radius of user cluster \\
        \hline
        {$\xi_i$}& {Arrival rate of packets intended for user $x_i$}\\
        \hline
        {$\beta$}& Mean delay requirement \\
        \hline
        {$h_0$}& Fading coefficient of desired link \\
        \hline
        {$g_i$}& Fading coefficient of interference link from BS $y_i$ \\
        \hline
        {$l_0$}& Distance between typical user and its associated BS \\
        \hline
        {$L_i$}& Distance between typical user and interfering BS $y_i$ \\
        \hline
        {$\alpha$}& Path loss exponent \\
        \hline
        {$\theta$}& Threshold of SIR \\
        \hline
        {$\mathbf{P}_s$}& Success probability \\
        \hline
        {$\delta$}& $\alpha$-related constant, defined as $\delta=2/\alpha$ \\
        \hline
        {$\tau$}& {Throughput of the typical BS }\\
        \hline
    \end{tabular}
\end{table}

The remaining part of the paper is organized as follows. In Section \ref{sec2}, we describe the system models. Then, we study the statistics of traffic in Section \ref{sec3} and evaluate the effect of traffic on network performance in Section \ref{sec4}. Numerical results are given in Section \ref{sec6}. Finally, Section \ref{sec7} concludes the paper. The notations are listed in Table \ref{notation_definition}.

\section{System Model}
\label{sec2}
\subsection{Network Structure}
\begin{figure}
\centering
\includegraphics[width=1\columnwidth]{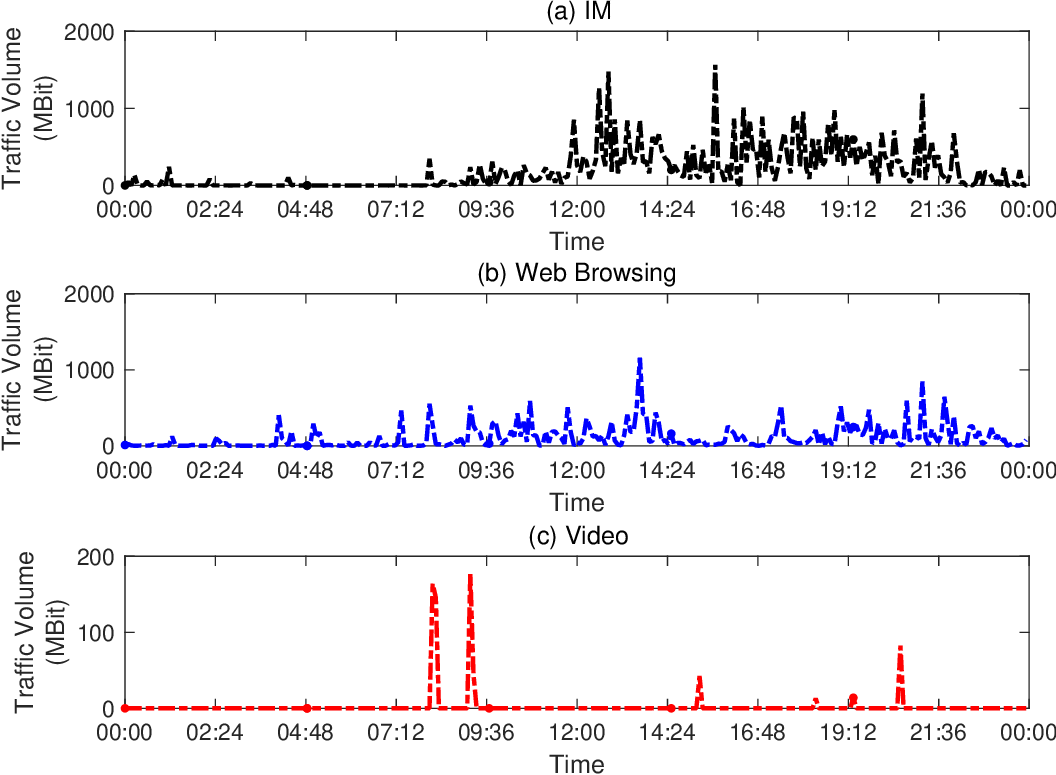} 
\caption{Cellular traffic of three different applications during one day in several randomly selected cells.}
\label{temporal}
\end{figure}

\begin{figure*}[!ht]
\centering
\includegraphics[width=2.1\columnwidth]{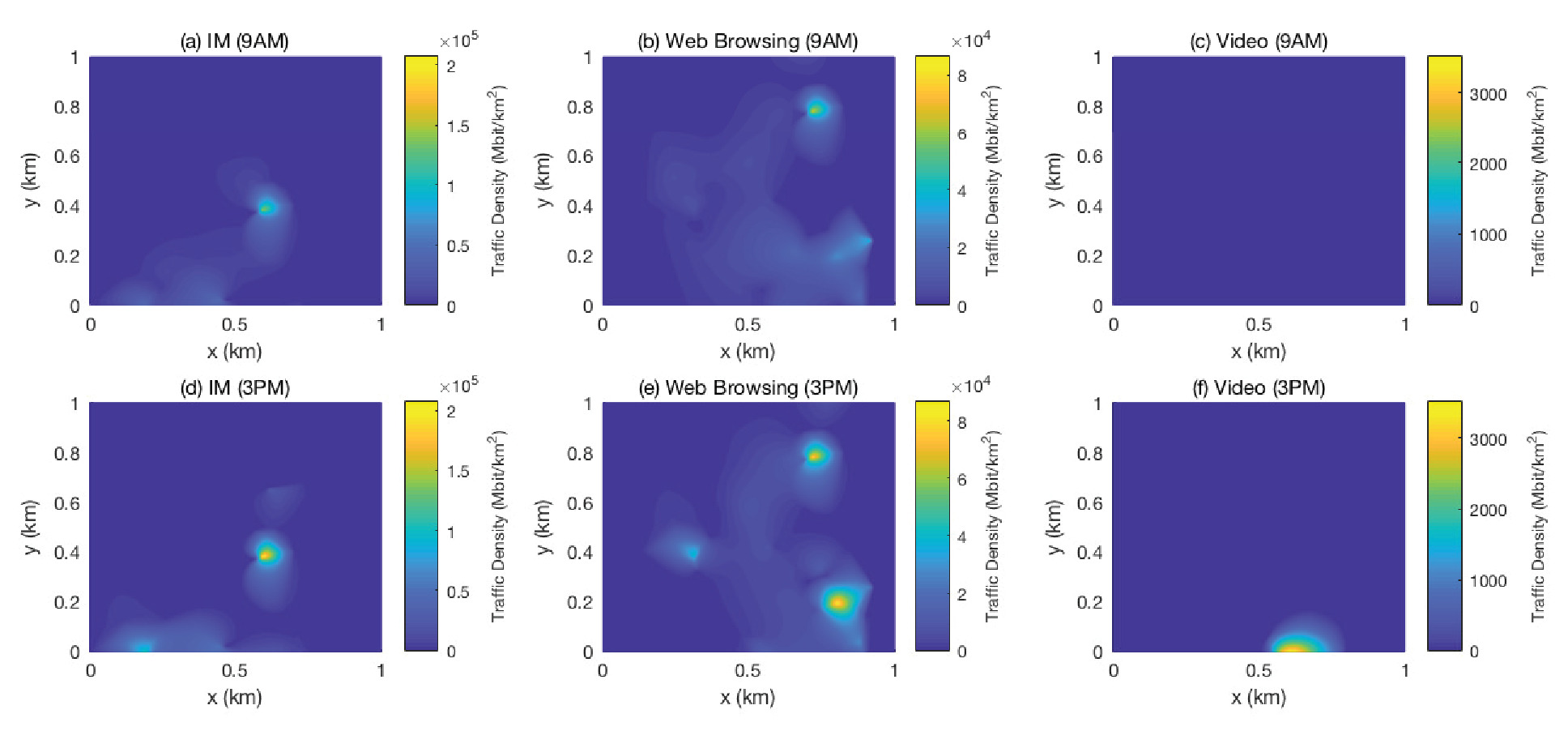} 
\caption{Heat map for cellular traffic of three wireless applications, i.e., instant messaging (IM), web browsing and video, at 9AM and 3PM in randomly selected dense urban areas. }
\label{spatialSparsity}
\end{figure*}

We first show the spatial and temporal distribution of the wireless traffic in the real scenarios. The data sets are based on a large number of practical traffic records from China Mobile in Hangzhou, an eastern provincial capital in China via the Gb interface of 2G/3G cellular networks or S1 interface of 4G cellular networks \cite{6381046}.
Figure \ref{temporal} shows the traffic amount for  three different applications during one day in the randomly selected cells, which is consistent with the experiments in \cite{7890496}.
In Figure \ref{spatialSparsity}, we plot the traffic density of three kinds of typical applications, i.e., instant messaging (IM), web browsing and video, at 9AM and 3PM in randomly selected dense urban areas. As shown in Figure \ref{spatialSparsity}, there are some ``hot spots'' varying in both temporal and spatial domain, and the spatially clustering property is also consistent with the experiments in \cite{7890496}.
Based on the characteristics of the real traffic in these figures, we propose an analytical model and study the effect of the traffic on the performance of wireless networks.

\begin{figure*}[!ht]
\centering
\includegraphics[width=1.4\columnwidth]{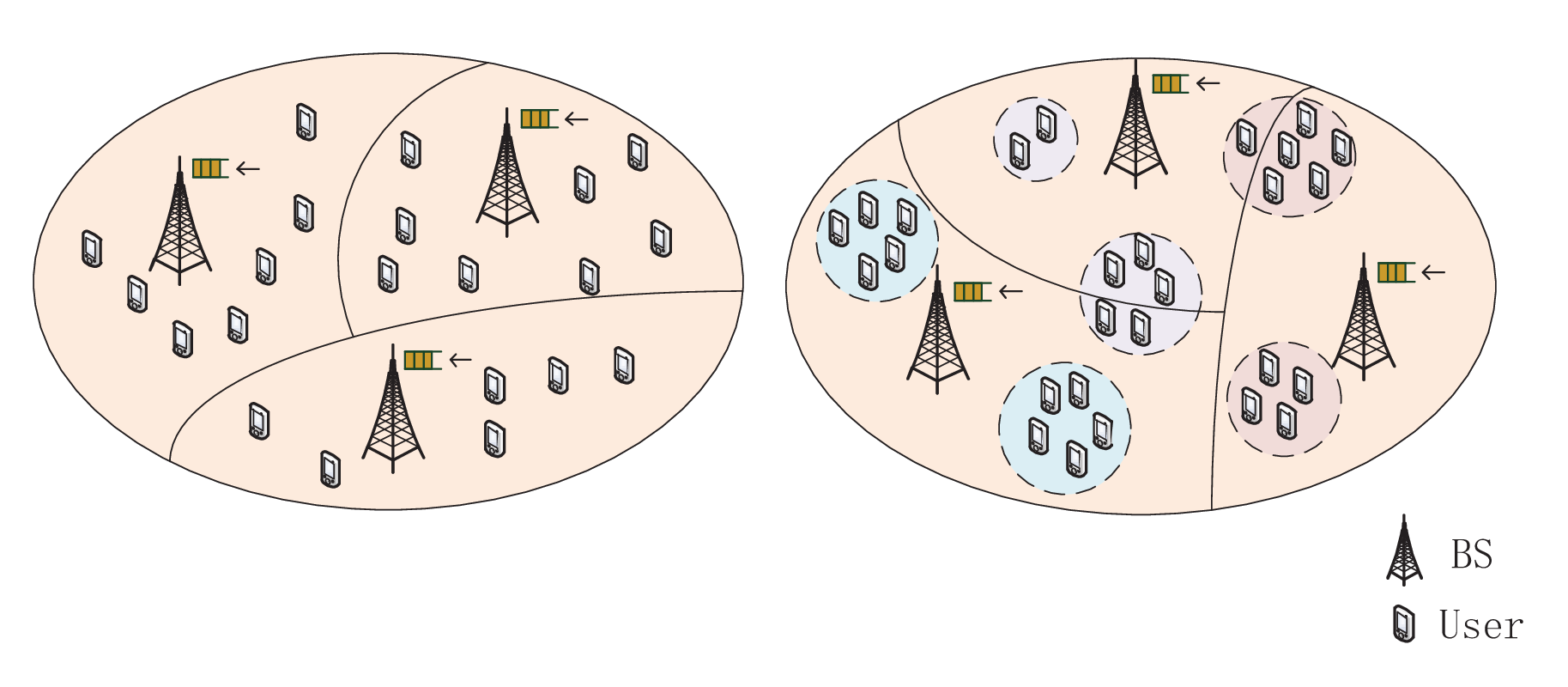} 
\caption{Illustration of the network model with BSs and users. Uniformly distributed traffic is modeled by the PPP in the left graph and non-uniformly distributed traffic is modeled by the PCP in the right graph.}
\label{fig:Network}
\end{figure*}

We consider a wireless network that consists of one tier of BSs and one tier of users, as illustrated in Figure \ref{fig:Network}. In this paper, we discuss the downlink of the wireless network with ALOHA channel access. The locations of BSs are modeled as a homogeneous PPP ${\Phi _b}= \left\{ {{y_i}} \right\}$ with intensity ${\lambda _b}$, denoted by ${\Phi _b} \sim {\rm{PPP}}({\lambda _b})$. The users are modeled by another point process ${\Phi _u} = \left\{ {{x_i}} \right\}$. The transmission power of the BSs is ${P_b}$. Since the network considered in the work is a single tier network, the shared spectrum policy is reasonable. The users in the network are shared with the same spectrum resource. In the time domain, the time is slotted into discrete time slots and the users will acquire the dominant right of the time slot with a random probability at each time slot.
We consider the following two kinds of spatial distributions for the users.
\begin{itemize}
\item{Case 1:} The locations of users form a homogeneous PPP of intensity ${\lambda _u}$ as shown in the left graph in Figure \ref{fig:Network}.
\item{Case 2:} The locations of users are distributed as a PCP as shown in the right graph in Figure \ref{fig:Network}. The centers of the clusters, i.e., the parent points, are distributed according to a PPP ${\Phi _p}$ of intensity ${\lambda _p}$. The users are uniformly scattered according to an independent PPP ${\Phi _x}$ of intensity ${\lambda _c}$ in the circular covered area of radius ${r_c}$ centered at each parent point $x \in {\Phi _p}$, which are called the daughter points. Thus, the distribution of all users is
\begin{equation}
{\Phi _u} = \bigcup\limits_{x \in {\Phi _p}} {{\Phi _x}}.
\end{equation}
In this case, the number of users in the typical cluster is a Poisson random variable with parameter $\pi r_c^2{\lambda _c}$, and the intensity of all users is ${\lambda _u} = \pi r_c^2{\lambda _c}{\lambda _p}$.
\end{itemize}

Each user is associated with the BS that provides the maximum average received power. Since the transmission power of all BSs is the same, each user will connect to the nearest BS. Without loss of generality, we consider a typical user located at the origin, and the typical user is associated with a BS located at $y_0$, which is named the typical BS.

\subsection{Packet Arrival}
Note that we consider the downlink of a wireless network. The packets intended for each user arrive at the associated BS and wait to be scheduled for delivering. Since each BS serves multiple users, we could consider that each BS maintains a separated queue (with infinite size) for each user to store the incoming packets. The number of queues at each BS equals to the number of users within the coverage of the BS. As a frequently-used packet arrival model in the discrete-time system, we assume that the packet arrival process intended for each user $x_i \in \Phi_u$ is a Bernoulli process \cite{21216,7886285} with arrival rate $\xi_i$. To characterize the diversity of traffic, we assume that the packet arrival rates $\{\xi_i\}$ are different for different users. In particular, we consider two cases where $\{\xi_i\}$ are i.i.d. exponentially distributed random variables and uniformly distributed random variables with mean $\mathbb{E}[\xi_i]=\xi_0$, respectively. According to the definition of Bernoulli process, the arrival rate is also the probability that a packet of the user ${x_i}$ arrives at the BS. The arrival processes for different users are assumed to be independent. The size of each packet is assumed to be fixed, and a BS requires exactly one time slot to deliver one packet to a user. If a packet fails for the transmission in the current time slot, it will be added into the head of queue and waits to be rescheduled.

\subsection{Channel and Scheduling}
\label{sec2_B}
Considering the propagation loss, we assume that the links between the serving BS and the users experience Rayleigh fading with unit mean. Then, the received signal power of a user at a distance $l$ from its serving BS is ${P_b}h{l^{ - \alpha }}$ where $h \sim {\rm{Exp}}(1)$. In other words, the power fading coefficients in different time slots are i.i.d. random variables and keep constant during one time slot. The fading coefficients of interference links are denoted by $\left\{{g_i}\right\}$. Since the noise is not the most focused and interested factor of our work and the network considered in the study consists of one tier of BSs and one tier of users where the interference from interfering BSs are more influential than the noise on the network performance, we ignore the noise power in the analysis. If the signal-to-interference ratio (SIR) at a user is larger than a constant threshold $\theta $, the receiver can successfully decode the packets. Otherwise, the packet will be failed for decoding, and the failed packets will be added into the head of the queues and wait to be scheduled again.

As for the scheduling mechanism of users at each BS, we consider the random scheduling of all active users whose queues are non-empty, i.e., each BS randomly selects a user to transmit from all users with non-empty queues in each time slot. Comparing with the scheduling scheme of randomly allocating time slots to all users in the coverage of a BS, the random scheduling of all active users is efficient without wasting the time slot resources.

\section{Traffic Statistic}
\label{sec3}
In order to evaluate the spatial and temporal fluctuation of the traffic, we consider two metrics, i.e., the probability distribution of the number of users served by a BS and the variance of total arrival rate. We first introduce the following lemma.
\begin{lemma}
\label{lemma1}
The PDF of the coverage area ${S}$ of a BS is approximated as \cite{6516167,ferenc2007size}
\begin{equation}
{f_{{S}}}\left( x \right) \simeq \frac{{343}}{{15}}\sqrt {\frac{{3.5}}{\pi }} {\left( {x{\lambda _b}} \right)^{2.5}}\exp \left( { - {3.5x{\lambda _b}}} \right){{\lambda _b}}. \label{equation6}
\end{equation}
\end{lemma}
\begin{IEEEproof}
In\cite{6516167}, the authors give the approximate PDF of the size of a macro-cell coverage area and the detailed mathematical derivations and simulations can refer to\cite{ferenc2007size}.
\end{IEEEproof}

Firstly, we derive the probability mass function (PMF) of the number of users for the uniformly distributed traffic where the users are modeled by a homogeneous PPP. Let $N$ be the number of users in a cell with given area $S$.
\begin{lemma}
\label{lemma3}
When users are distributed as a PPP and associated with BSs that providing maximum average received power, the PMF of $N$ is \cite{haenggi2012stochastic}
\begin{equation}
\mathbb{P}\{N=k\} = \frac{e^{-\lambda_uS}}{k!}{(\lambda_uS)^k}. \label{equation13}
\end{equation}
\end{lemma}
\begin{IEEEproof}
The PMF of $N$ is obtained by the definition of two-dimensional PPP and the author gives detailed mathematical derivations and descriptions of general PPP in \cite[ch.2] {haenggi2012stochastic}.
\end{IEEEproof}

In order to derive the PMF of $N$ for the non-uniformly distributed case, we introduce a new association rule where the users will access to the BS nearest to the parent point. Since we assume that the users access to the nearest BSs in the previous section, we give the following Approximation \ref{appro_associationRules} to present the relationships between the two association rules.
\begin{approximation}
In the PCP case, when the radius of cluster is much smaller than the radius of cell, the association rule that users access to the nearest BS can be regarded as the association rule that the users access to the BS nearest to the parent point.
\label{appro_associationRules}
\end{approximation}

\begin{figure}[!ht]
\centering
\includegraphics[width=1\columnwidth]{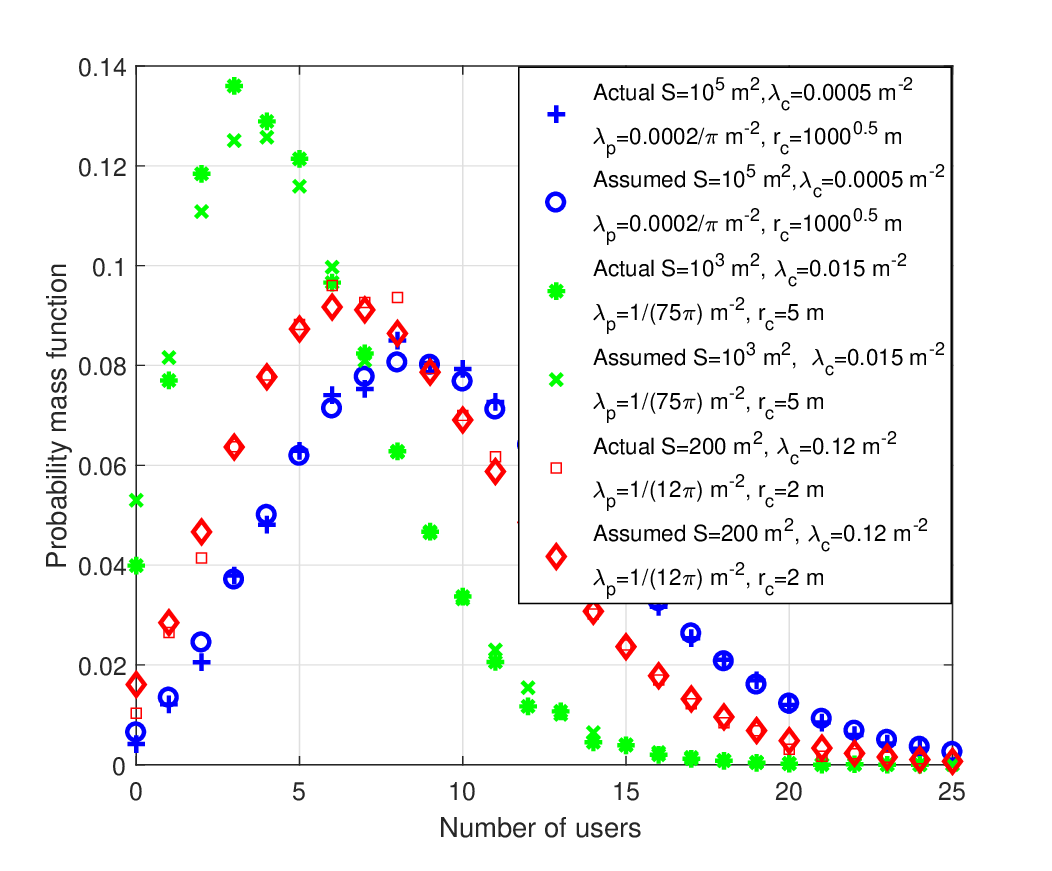} 
\caption{The PMF of $N$ when the users adopt different association rules.}
\label{fig:PMF_N_associationRule_Sim_PCP}
\end{figure}

\begin{figure}
  \centering
  \subfigure[Small cell size]{
    \label{N_lambdaP_smallS} 
    \includegraphics[width=3.2in]{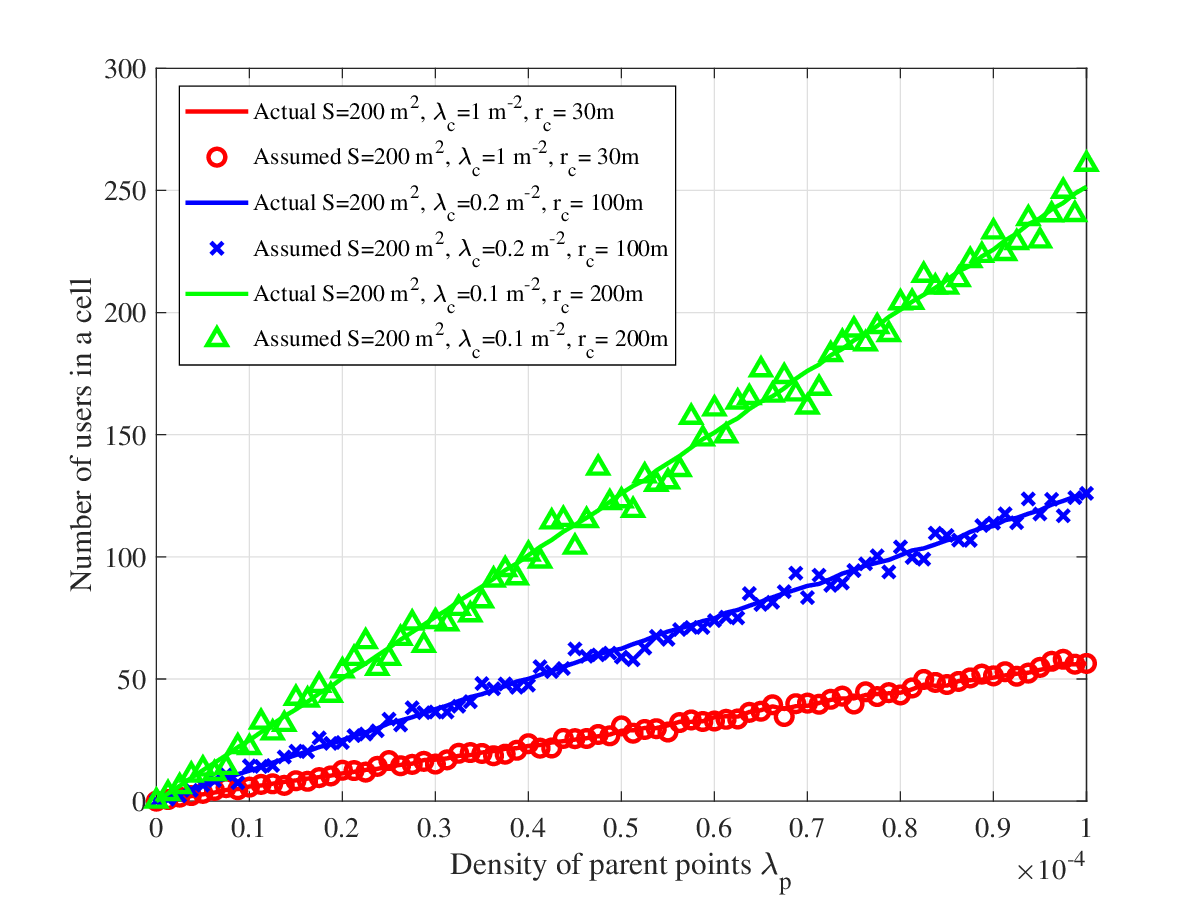}}
  \hspace{1in}
  \subfigure[Large cell size]{
    \label{N_densityP_largeS} 
    \includegraphics[width=3.2in]{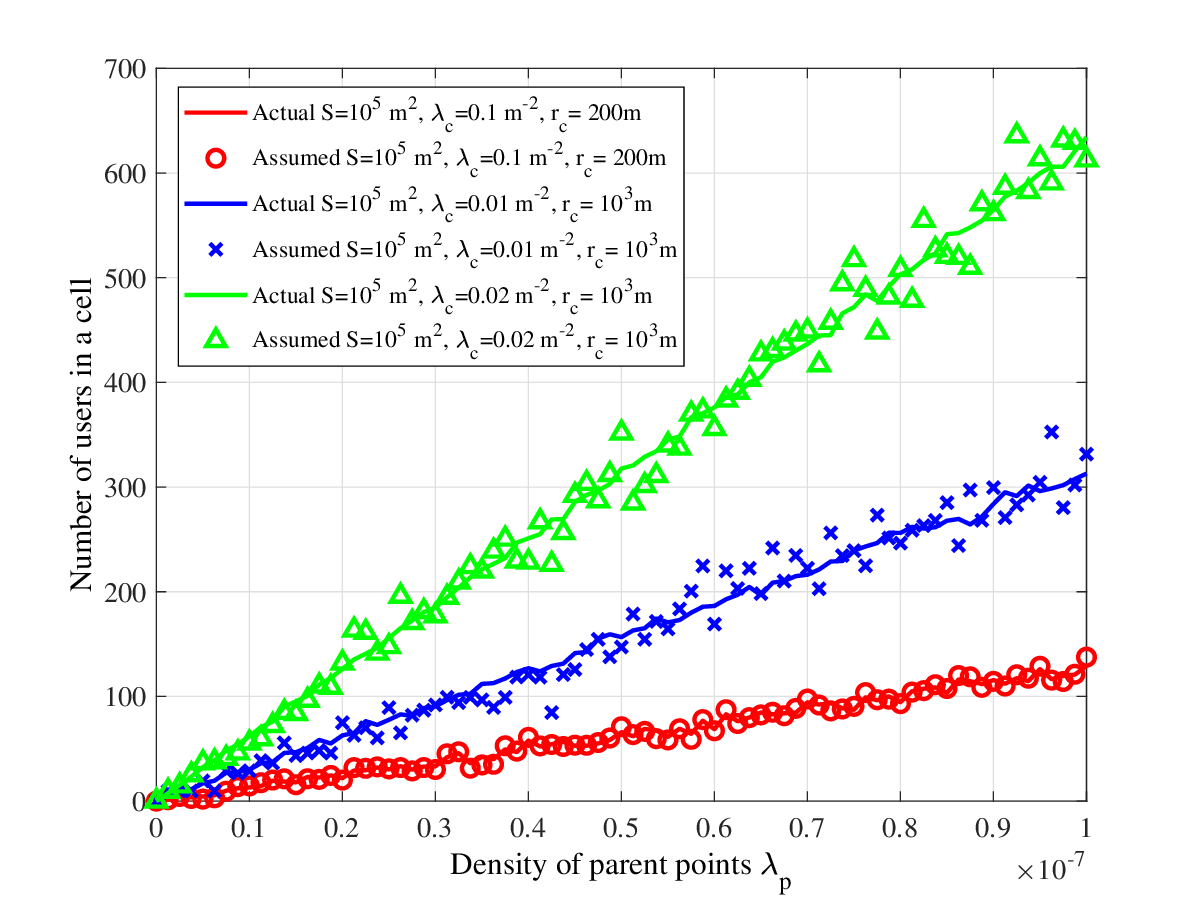}}
  \caption{Number of users in a cell as functions of $\lambda _p$ when the users form a PCP with given area $S$.}
  \label{N_densityP_smallS_largeS} 
\end{figure}

In order to validate this approximation, we give the simulation results Figure \ref{fig:PMF_N_associationRule_Sim_PCP} and Figure \ref{N_densityP_smallS_largeS}. The curves `Actual' denote the case where each user accesses the nearest BS. The curves `Assumed' denote the case where the users access to the BS nearest to the parent point. In Figure 4, we compares the PMFs when considering different association rules. In particular, we plot the `Actual' and `Assumed' curves for different cell areas $S$ and various PCP parameters to verify the accuracy of the approximation. It is observed that the `Assumed' curve approaches to the `Actual' curve, indicating that the approximation is reasonable. Based on Approximation 1, we obtain the following lemma.

\begin{lemma}
When users are distributed as a PCP and associated with BSs that providing maximum average received power, the PMF of $N$ is
\begin{align}
\mathbb{P}\{ N = k \} &\approx \sum\limits_{a = 0}^\infty  {\mathbb{P}\{ {{N_p} = a,{N_d} = k} \}} \nonumber \\
&= \sum\limits_{a = 0}^\infty  {\frac{{{e^{ - {\lambda _p}S}}}}{{a!}}} {\left( {{\lambda _p}S{e^{ - {\lambda _c}\pi r_c^2}}} \right)^a}\frac{{{{\left( {{\lambda _c}a\pi r_c^2} \right)}^k}}}{{k!}},
\label{equation14}
\end{align}
where $N_p$ is the number of parent points, and $N_d$ is the number of daughter points.
\label{lemma4}
\end{lemma}

\begin{IEEEproof}
When users access the BS nearest to the parent point, the PMF of $N$ is
\begin{align}
\mathbb{P}\left( {N = k} \right) &\stackrel{(a)}{=} \sum\limits_{a = 0}^\infty \mathbb{P}\{N_p=a\}\mathbb{P}\{ N_d = k| {N_p = a} \}
\nonumber \\
&= \sum\limits_{a = 0}^\infty  {\frac{{{e^{ - {\lambda _p}S}}}}{{a!}}} {\left( {{\lambda _p}S{e^{ - {\lambda _c}\pi r_c^2}}} \right)^a}\frac{{{{\left( {{\lambda _c}a\pi r_c^2} \right)}^k}}}{{k!}},
\end{align}
where (a) follows from the total probability formula. Based on the Approximation 1, we obtain Lemma 3.
\end{IEEEproof}

\begin{figure}[!ht]
\centering
\includegraphics[width=1\columnwidth]{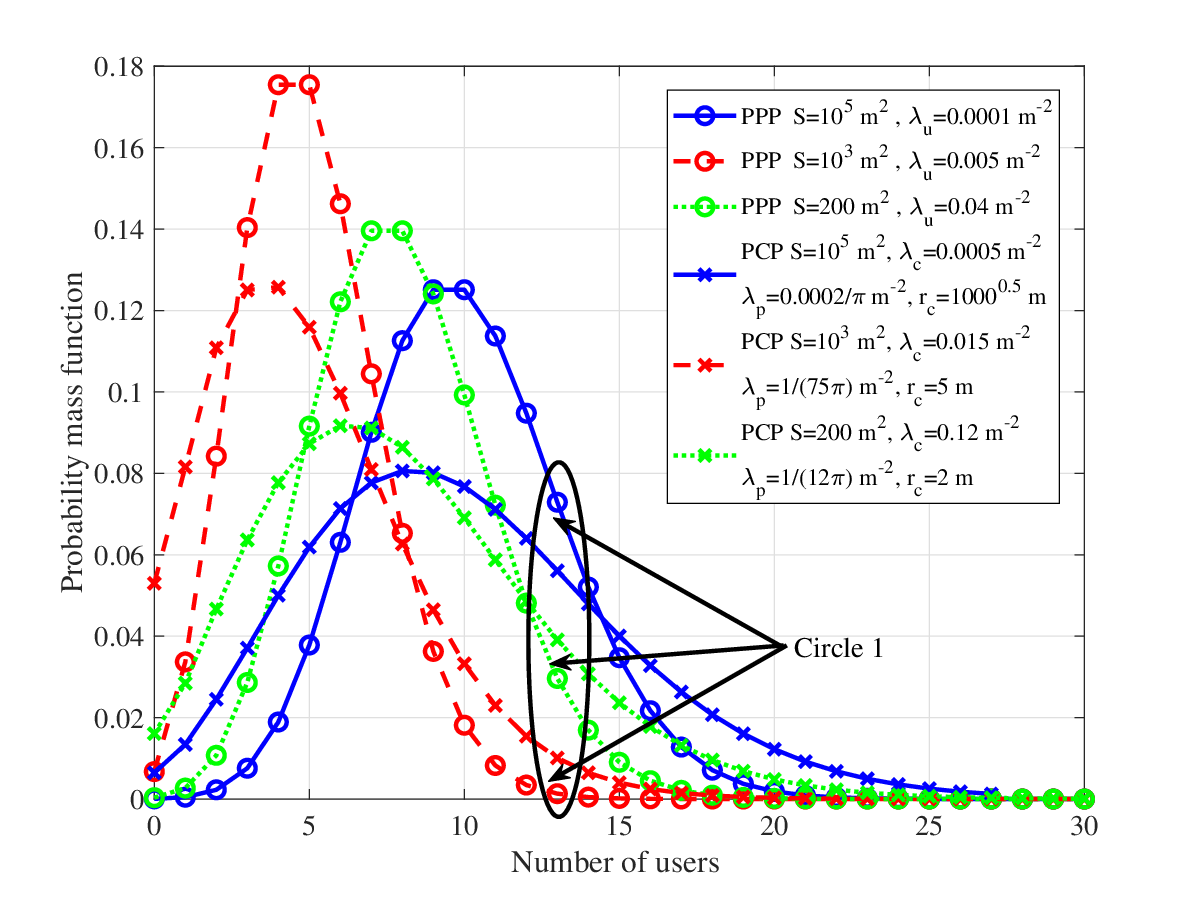} 
\caption{The PMF of the number of users when the users form a PPP or a PCP with the same $\lambda _u$ and given area $S$.}
\label{fig:PMF_N_PC}
\end{figure}

In Figure \ref{fig:PMF_N_PC}, we plot the PMF of the number of users in the cases where the users form a PPP or a PCP with the same $\lambda _u$ in a cell with area $S$. We observe that, with given cell area $S$, the probability that the number of users is either very small or very large in the PCP case is larger than that in the PPP case. This is partly attributed to the fact that the users in the PCP case appear to be more grouped.
Meanwhile, the PMF in the PPP case is more centralized so the probability is larger than that in the PCP case when the number of users is a medium value. In both cases, as the value of ${\lambda_u}S$ increases gradually, the peak value of the PMF decreases since the PMF is more dispersed when the mean number of users increases. As shown in circle 1, the number of users increases due to the increase of the mean number of users ${\lambda_u}S$.

Let ${\xi _{j,\rm{total}}}$ be the total arrival rate of packets in the coverage area $S_j$ of a BS $y_j$, we get
\begin{align}
{\xi _{j,\rm{total}}} = \sum\limits_{x_i \in \Phi_{u,S_j}}{\xi _i },
\end{align}
where $\Phi_{u,S_j}$ is the set of users in area $S_j$. $N_j$ is the total number of users in area $S_j$. The mean total arrival rate in area $S_j$, $\mathbb{E}[ \xi _{j,\rm{total}} ] = {\mathbb{E}[\xi_i]}\mathbb{E}[N_j]$, is
\begin{align}
\mathbb{E}[\xi _{j,\rm{total}}] = \mathbb{E}[\xi_i]{\mathbb{E}_{S}}\big[{\mathbb{E}\left[ {N_j\left| S \right.} \right]} \big]
= \mathbb{E}[\xi_i]\frac{\lambda_u}{\lambda_b}.
\label{equation17}
\end{align}
The variance of the total arrival rate in the area $S_j$ is $\mathbb{D}[ {{\xi _{j,\rm{total}}}} ] = \mathbb{E}[ {\xi_{j,\rm{total}}^2}] - {\big( {\mathbb{E}[ {{\xi _{j,\rm{total}}}} ]} \big)^2}$.

Let ${\mathrm{D}_{P,\xi _{j,\rm{total}}}}$ be the variance of total arrival rate in the uniformly distributed case, and ${\mathrm{D}_{C,\xi _{j,\rm{total}}}}$ be the variance of total arrival rate in the non-uniformly distributed case.
With the above obtained mean total arrival rate, we obtain the following lemma.
\begin{lemma}
When users are associated with BSs that providing maximum average received power, the variance of total arrival rate in the uniformly distributed case is
\begin{align}
{\mathrm{D}_{P,\xi _{j,\rm{total}}}} \approx {(\mathbb{E}[\xi_i])^2}\left( {0.2857\frac{{\lambda _u^2}}{{\lambda _b^2}} + \frac{{{\lambda _u}}}{{{\lambda _b}}}} \right).
\end{align}
Meanwhile, the variance in non-uniformly distributed case is
\begin{align}
{\mathrm{D}_{C,\xi _{j,\rm{total}}}} \approx {({\mathbb{E}[\xi_i]})^2}
 \left( {0.2857\frac{{\lambda _u^2}}{{\lambda _b^2}} + \frac{{{\lambda _u}}}{{{\lambda _b}}}( {\pi r_c^2{\lambda _c} + 1} )} \right).
\end{align}
\end{lemma}

\begin{IEEEproof}
When the users are distributed as a PPP, the variance of the total arrival rate is
\begin{align}
{\mathrm{D}_{P,\xi _{j,\rm{total}}}} &= \mathbb{E}\left[ {{{\xi _{j,\rm{total}}^2}}} \right] - {\big( {\mathbb{E}\left[ {{\xi _{j,\rm{total}}}} \right]} \big)^2}
\nonumber \\
&= {\mathbb{E}_{{S_j}}}[ {{\mathrm{E}_{P,\xi _{j,\rm{total}}^2}}} ] - {\left( \mathbb{E}[\xi_i]\frac{{{\lambda _u}}}{{{\lambda _b}}} \right)^2}.
\label{equation20}
\end{align}

In \eqref{equation20}, the mean ${\mathrm{E}_{P,\xi _{j,\rm{total}}^2}}$ is
\begin{align}
{\mathrm{E}_{P,\xi _{j,\rm{total}}^2}} &= \mathbb{E}[ {{{\left( {\mathbb{E}[\xi_i]} \right)}^2}N_j^2} ]
\nonumber \\
                                       &= {\left( {\mathbb{E}\left[ \xi_i \right]} \right)^2}\sum\limits_{k = 0}^\infty  {{k^2}\mathbb{P}\{N_j=k\}}
\nonumber \\
                                       &\stackrel{(a)}{=} {\left( {\mathbb{E}[\xi_i]}\right)^2}\left( {\lambda _u^2{S^2} + {\lambda_u}S} \right),
\label{equation21}
\end{align}
where (a) follows from the result
\begin{align}
\sum\limits_{k = 0}^\infty  {k^2}\mathbb{P}\{ {{N_j} = k} \} &= \sum\limits_{k = 0}^\infty  {{k^2}\frac{{{e^{ - {\lambda _u}S}}}}{{k!}}{{\left( {{\lambda _u}S} \right)}^k}}
\nonumber \\
&= {e^{ - {\lambda _u}S}}{\lambda _u}S\sum\limits_{k = 0}^\infty  {k\frac{{{{\left( {{\lambda _u}S} \right)}^{k - 1}}}}{{\left( {k - 1} \right)!}}}
\nonumber \\
&=  {\lambda _u^2{S^2} + {\lambda _u}S}.
\end{align}

Plugging \eqref{equation21} into the equation \eqref{equation20}, we get the variance of the total arrival rate as
\begin{align}
{\mathrm{D}_{P,\xi _{j,\rm{total}}}}
&\stackrel{(a)}{=} {\big( {\mathbb{E}[\xi_i]} \big)^2}\left( {\int_0^\infty  {\left( {\lambda _u^2{x^2} + {\lambda _u}x} \right){f_{S_j}}(x)dx} } \right.\left. { - \frac{{\lambda _u^2}}{{\lambda _b^2}}} \right)
\nonumber \\
&\stackrel{(b)}{\approx} {\big( {\mathbb{E}[ \xi_i ]} \big)^2}\left( {0.2857\frac{{\lambda _u^2}}{{\lambda _b^2}} + \frac{{{\lambda _u}}}{{{\lambda _b}}}} \right),
\label{equation22}
\end{align}
where $(a)$ follows from the PDF of $S_j$, and $(b)$ follows from the definition of $f_{{S_j}}(\cdot)$ given by \eqref{equation6} and the calculation of the integral as follows.
\begin{align}
&\int_{\rm{0}}^\infty  {\left( {\lambda _u^2{x^2} + {\lambda _u}x} \right){f_{{S_j}}}\left( x \right)dx} \nonumber \\
& = \int_{\rm{0}}^\infty  {\left( {\lambda _u^2{x^2} + {\lambda _u}x} \right)\frac{{343}}{{15}}\sqrt {\frac{{3.5}}{\pi }} {{\left( {x{\lambda _b}} \right)}^{2.5}}{e^{ - 3.5{\lambda _b}x}}{\lambda _b}dx}
\nonumber \\
& \stackrel{(a)}{=} \frac{{343}}{{15}}\sqrt {\frac{{3.5}}{\pi }} \lambda _b^{3.5}\left( {\lambda _u^2\frac{{\Gamma \left( {5.5} \right)}}{{{{\left( {3.5{\lambda _b}} \right)}^{5.5}}}} + {\lambda _u}\frac{{\Gamma \left( {4.5} \right)}}{{{{\left( {3.5{\lambda _b}} \right)}^{4.5}}}}} \right)
\nonumber \\
& \approx 1.2857\frac{{\lambda _u^2}}{{\lambda _b^2}} + \frac{{{\lambda _u}}}{{{\lambda _b}}},
\label{integral_1}
\end{align}
where (a) follows from the integral $\int_{\rm{0}}^\infty  {{x^m}{e^{ - \beta {x^n}}}dx}  = \frac{{\Gamma \left( r \right)}}{{n{\beta ^r}}},r = \frac{{m + 1}}{n}$. $\Gamma \left( x \right) = \int_0^\infty  {{t^{x - 1}}{e^{ - t}}dt}$ denotes the standard gamma function.

When the users are distributed as a PCP, the variance of the total arrival rate is
\begin{align}
{\mathrm{D}_{C,\xi _{j,\rm{total}}}} = {\mathbb{E}_{{S_j}}}\big[ {{\mathrm{E}_{C,\xi_{j,\rm{total}}^2}}} \big] - {\left( {\mathbb{E}[\xi_i]\frac{{{\lambda _u}}}{{{\lambda _b}}}} \right)^2}.
\label{equation23}
\end{align}

In the above equation \eqref{equation23}, the mean ${{\mathrm{E}_{C,\xi _{j,\rm{total}}^2}}}$ is
\begin{align}
{\mathrm{E}_{C,\xi _{j,\rm{total}}^2}} = {\left( {\mathbb{E}[ \xi_i ]} \right)^2}{\mathbb{E}}[{N_j}^2]
= {\left( {\mathbb{E}[\xi_i]} \right)^2}( {{\mathbb{D}}[N_j] + {{( {{\mathbb{E}}[ N_j ]} )}^2}} ),
\label{equation24}
\end{align}
where ${N_j} = \sum\limits_{i = 1}^{{N_p}} {{N_{{C_i}}}} $ is a Compound Poisson random variable and ${N_{{C_i}}} = {N_{{d_i}}}$ is the number of users in the $i$th cluster. According to the properties of Compound Poisson random variable, we have ${\mathbb{E}}[ {{N_j}} ] = \mathbb{E}[ {{N_p}} ]\mathbb{E}[ {{N_C}} ]$ and ${\mathbb{D}}[ {{N_j}} ] = \mathbb{E}[ {{N_p}} ]( {\mathbb{D}[ {{N_C}} ] + {{( {\mathbb{E}[ {{N_C}} ]} )}^2}} )$. $N_p$ and $N_C$ are Poisson random variable with mean $S{\lambda _p}$ and $\pi r_c^2{\lambda _c}$, respectively. Therefore, the mean and variance of $N_C$ is $\mathbb{D}[ {{N_C}} ] = \mathbb{E}[ {{N_C}} ] = \pi r_c^2{\lambda _c}$, and the mean of $N_p$ is $\mathbb{E}[ {{N_p}} ] = S{\lambda _p}$.

Then, the equation \eqref{equation24} can be derived as
\begin{align}
{\mathrm{E}_{C,\xi _{j,\rm{total}}^2}} &= {( {\mathbb{E}[ \xi_i ]} )^2}\Big( {\mathbb{E}[ {{N_p}} ]\mathbb{D}[ {{N_C}} ] + {\mathbb{E}[ {{N_p}} ]{{( {\mathbb{E}[ {{N_C}} ]} )}^2} }}  \Big. \nonumber \\
 & \Big. + {{( {\mathbb{E}[ {{N_p}} ]\mathbb{E}[ {{N_C}} ]} )}^2} \Big).
\label{equation25}
\end{align}
Similar to that in \eqref{equation22}, we have
\begin{align}
{\mathrm{D}_{C,\xi _{j,\rm{total}}}} \approx {\left( {\mathbb{E}\left[ \xi_i \right]} \right)^2}  \left( {0.2857\frac{{\lambda _u^2}}{{\lambda _b^2}} + \frac{{{\lambda _u}}}{{{\lambda _b}}}\left( {\pi r_c^2{\lambda _c} + 1} \right)} \right).
\end{align}
\end{IEEEproof}

\begin{figure}[!ht]
\centering
\includegraphics[width=3.5in]{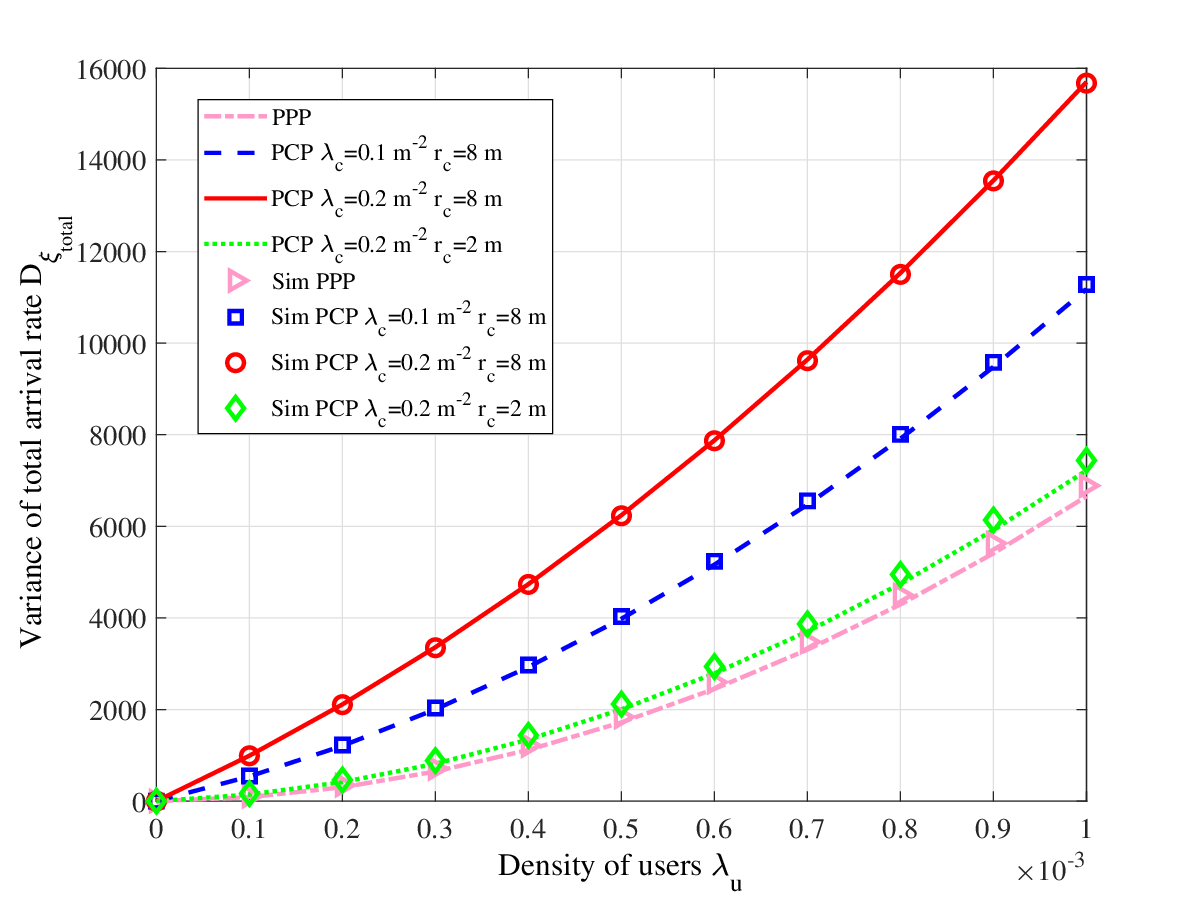} 
\caption{Variances of the total arrival rate in a given area $S$ for the uniformly and non-uniformly distributed cases ($\lambda _b=0.00001 m^{-2} , \mathbb{E}\left[ \xi_i \right]=1.5$).}
\label{D_natu_cal_sim}
\end{figure}

In Figure \ref{D_natu_cal_sim}, we plot the variances of total arrival rate in a given area $S$ for the uniformly and non-uniformly cases. We observe that the variance of total arrival rate increases when $\lambda_u$ increases. This is because when the number of users increases, the difference of the arrival rate among the users will be greater. In particular, we observe that the variance of total arrival rate is larger in the non-uniformly distributed case than that in the uniformly distributed case. We also compare the variance in the non-uniformly distributed case for different $r_c$ and $\lambda _c$. When fixing the radius of cluster $r_c$, the variance of the total arrival rate increases with the increase of the density of the clusters $\lambda _c$. When fixing the density of the clusters $\lambda _c$, the variance of the total arrival rate increases with the increase of the radius of cluster $r_c$. The reason is that when increasing the number of users in a cluster, the difference of the arrival rates also increases.

\section{Performance Evaluation}
\label{sec4}
In this section, we explore the effect of traffic on the network performance from the aspects of success probability, throughput, delay and stability. As mentioned in the system model, we consider a typical user at the origin, and the serving BS of the typical user is located at $y_0$. Let $l_0$ be the distance between the typical user at the origin and the nearest BS $y_0$. The PDF of $l_0$ can be derived according to the fact that the null probability of a 2-D Poisson process with intensity $\lambda_b$ in an area $A$ is $e^{ - {\lambda}_b A}$. Thus, the PDF of $l_0$ is \cite{6042301}
\begin{align}
{f_{l_0}}(l_0) = 2\pi \lambda_b l_0 e^{- \lambda_b \pi l_0^2}.
\label{equation27}
\end{align}
The distance $l_0$ is a random variable. A similar model is described by a meta distribution in \cite{8168355}.

Note that we consider the downlink of a wireless network, and the status of a BS may be either busy or idle.
Letting $L_i$ be the distance between the typical user and the interfering BS $y_i$, the SIR at the typical user with a distance $l_0$ from its serving BS in the time slot $t$ is
\begin{align}
{{\rm{SIR}}_t} = \frac{{{P_b}{h_0}{l_0^{ - \alpha }}}}{{\sum\limits_{{y_i} \in {\Phi _b}\backslash {y_0}} {{\mathbbm{l}_{ {{{y}_i}} }}{P_b}{g_i}L_i^{ - \alpha }} }},
\label{equation28}
\end{align}
where ${\mathbbm{l}_{y_i}}$ is the indicator function defined such that ${\mathbbm{l}_{y_i}} = 1$ holds if the interfering BS $y_i$ is active, while ${\mathbbm{l}_{y_i}}=0$ holds otherwise.
The denominator is the interference at the typical user given by
\begin{align}
I = \sum\limits_{{y_i} \in {\Phi _b}\backslash {y_0}} {{\mathbbm{l}_{{y_i}}}{P_b}{g_i}{L_i^{ - \alpha }}}.
\end{align}
When evaluating the expectation with respective to the BS deployment pattern $\Phi_b$, the distance $l_0$ should be regarded as a random variable with PDF given by (\ref{equation27}).

The interference is determined by the active BSs whose queues are non-empty in current time slot. Note that the statuses of queues at different BSs change over the time. In order to make the analysis of success probability feasible, we introduce a factor $q$ which characterize the busy probability (or the active probability) of all BSs on average in the network. In the following discussions, we first propose methods to determine the busy probability $q$ for the aforementioned two scheduling schemes, then we use the obtained busy probability to derive several performance metrics.

\subsection{Busy Probability}
For the random scheduling of active users, a BS is active as long as any queue of the users is non-empty. To derive the busy probability of a BS directly is rather difficult since the number of active users served by a BS varies with the time slots. However, in order to derive the busy probability of BSs, we could think about the problem from another point of view. We could consider that each BS maintains a large queue to store the incoming packets and do not distinguish which user the packets belong to. Then, the average arrival rate of the queueing process at each BS becomes $N\xi_0$, and the service rate of the large queue will be $\mathbf{P}_{s|\Phi_b}$, which is the conditional success probability given the realization of $\Phi _b$, defined as
\begin{align}
\mathbf{P}_{s|\Phi_b} \buildrel \Delta \over = \mathbb{P} \{\rm{SIR}_t > \theta \mid \Phi_b \}.
\end{align}

According to the property of the G/G/1 queueing system, conditioned on the realization of $\Phi_b$, the probability for the large queue at a BS being non-empty equals to the utilization of the queueing system at the large queue, which is
\begin{align}
\mathbf{P}_{a|\Phi_b}=\min\left\{\frac{N\xi_0}{{\mathbf{P}_{s\mid\Phi_b } }},1\right\}. \label{eqn:Paroundrobin}
\end{align}
Note that the condition for a BS being busy at certain time slot is equivalent to the condition that the large queue is non-empty. Thus, the above non-empty probability is also the busy probability of a BS conditioned on $\Phi_b$.

\begin{rem}
If we consider the random scheduling of all users other than just the active users in the coverage of a BS, the probability of a BS being busy equals the probability that the queue of a randomly scheduled user is non-empty. Since the arrival rate of packets intended for each user is $\xi_0$ while the service rate is $\mathbf{P}_{s\mid\Phi_b}/N$ due to the randomly scheduling, the busy probability will be the same as that given by the equation \eqref{eqn:Paroundrobin}. The result is exactly coincident with that for random scheduling of just active users, which is rather counterintuitive. It can be interpreted as that although the random scheduling of all users may allocate time slots to users with empty queues, the busy probability of BSs will not be reduced since it increases the probability that a queue being non-empty. With these discussions, we arrive the conclusion that the busy probability $q$ for random scheduling of active users equals to that for random scheduling of all users.
\end{rem}

\begin{figure}[!ht]
\centering
\includegraphics[width=1\columnwidth]{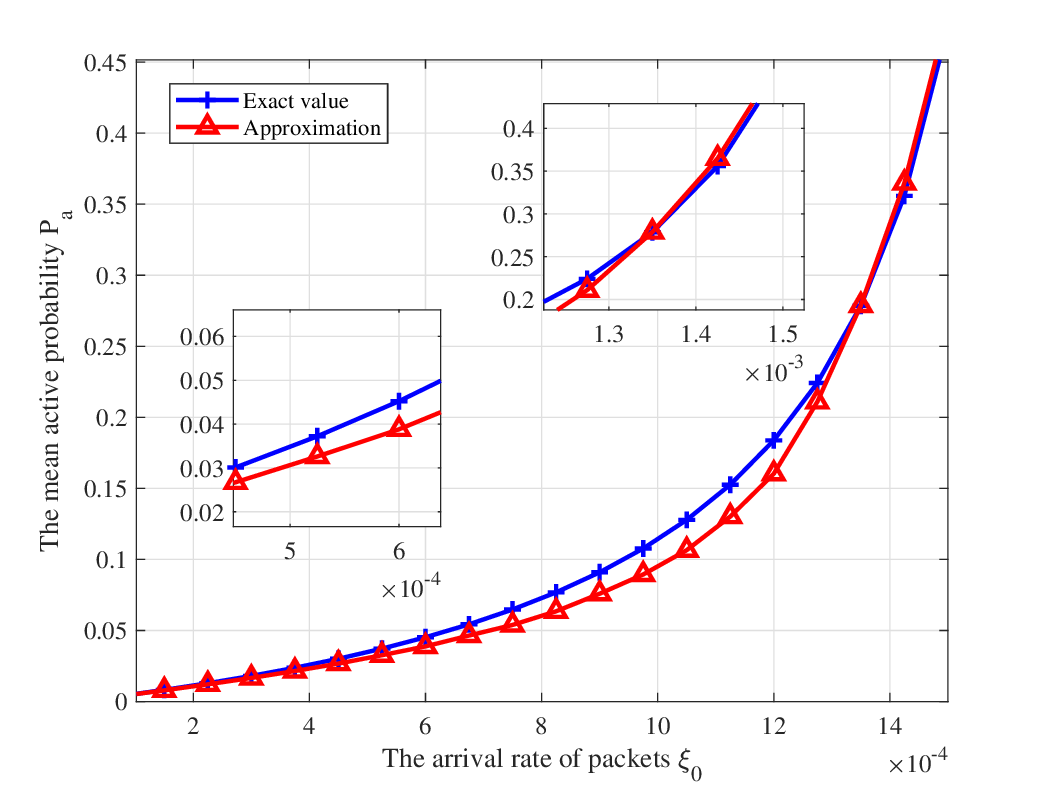}
\caption{Illustration of the accuracy for the approximation given by (\ref{equation26}). The parameters are set as $\theta=10, \lambda_b=10^{-5}, N = 50$ and $\alpha = 3$.}
\label{fig:Pa_left_right_eq26}
\end{figure}

By averaging over the point process $\Phi_b$, we get the mean active probability of all BSs as
\begin{align}
\mathbf{P}_a  = \mathbb{E}_{\Phi _b } \left[ {\min \left\{\frac{N\xi_0}{\mathbf{P}_{s\mid\Phi_b}},1 \right\}} \right] \stackrel{(a)}{\approx} \min \left\{\frac{N\xi_0}{\mathbf{P}_s},1\right\},
\label{equation26}
\end{align}
where $(a)$ holds by using $\min({1}/{\mathbb{E}[X]}, 1)$ to approximate $\mathbb{E}[\min(1/X, 1)]$. The accuracy of the approximation is demonstrated by Figure \ref{fig:Pa_left_right_eq26}, in which the blue curve with cross-shaped markers denotes the exact value of mean active probability, while the red curve with triangular markers plots the approximated mean active probability given by (\ref{equation26}).

Note that in the previous discussion we have assumed that the busy probability of BSs is $q$. The approximated mean active probability given by (\ref{equation26}) should also equal to $q$. Since the success probability $\mathbf{P}_{s}$ is a function of $q$, we get a fixed-point equation as
\begin{align}
\min \left\{ {\frac{N\xi_0}{\mathbf{P}_s},1} \right\} = q.
\label{approxeq31}
\end{align}
By solving the above fixed-point equation, we get the following lemma which gives the busy probability of BSs in the steady state for random scheduling of active users.

\begin{lemma}
\label{lemma6}
The solution of equation \eqref{approxeq31} is given by
\begin{align}
q = \left\{ \begin{aligned}
&\frac{ N{\xi _0}{\mathop{\rm sinc}\nolimits} (\delta)}{{{\mathop{\rm sinc}\nolimits} (\delta) - N{\xi_0}{\theta ^{\delta}}}}, & &{\mathrm{if} \  0 < \xi_0 < \frac{\mathrm{sinc}(\delta)}{N({\mathrm{sinc}(\delta) + \theta^\delta})}}\\
&1, & &{\mathrm{if} \  {\xi_0} \ge \frac{\mathrm{sinc}(\delta)}{N({\mathrm{sinc}(\delta) +\theta^\delta} )}}
\end{aligned} \right. .
\label{equation32}
\end{align}
where $\delta=2/\alpha$ and $\mathrm{sinc}(\cdot)$ is the sinc function.
\end{lemma}
\begin{IEEEproof}
Combining \cite[eq.5.14] {haenggi2012stochastic} with \cite{6042301}, we get the success probability conditioned on the link distance $l_0$ as
\begin{align}
\mathbf{P}_{s\left| {l_0 } \right.}  = \exp \left( { - \frac{{\pi \lambda _b q\theta ^{\delta} l_0 ^2 }}{{{\mathop{\rm sinc}\nolimits} ( \delta)}}} \right),
\end{align}
where the interference is obtained by the independent thinning of the original Poisson point process with probability $q$. Averaging over $l_0$, we get the success probability as
\begin{align}
\mathbf{P}_s &= {\mathbb{E}_{l_0}}[\mathbf{P}_{s\mid l_0}] = \int_0^\infty {{\exp\left(-\frac{\pi\lambda_bq\theta ^\delta x^2}{\mathrm{sinc}(\delta)}\right)}f_{l_0}(x)\mathrm{d}x} \nonumber \\
&= \frac{\mathrm{sinc}(\delta)}{\mathrm{sinc}(\delta) + q\theta^\delta}.
\label{equation34}
\end{align}
Plugging $\mathbf{P}_s$ into \eqref{approxeq31}, we get the fixed-point equation \eqref{approxeq31} as
\begin{align}
\min\bigg\{\frac{N\xi_0(\mathrm{sinc}(\delta)+q\theta^\delta)}{\mathrm{sinc}(\delta)},1\bigg\} = q.
\label{equation}
\end{align}
In order to solve the above equation \eqref{equation}, we consider two cases, $0 < \frac{{N\xi_0( {\mathrm{sinc}(\delta) + q{\theta ^{\delta}}})}}{{\mathrm{sinc}(\delta)}} < 1$ and $\frac{{N{\xi _0}\left( {\mathrm{sinc}(\delta) + q{\theta ^{\delta}}} \right)}}{{\mathrm{sinc}(\delta)}} \ge 1$. In order to clarify the solving process for the fixed-point equation, we give the geometric explanation in the following. In Figure \ref{fig:xy_label1}, the red curve plots the solution for the equation \eqref{equation}, and the red dot indicates the intersection of the curve $q = \frac{{N\xi _0 \mathrm{sinc}(\delta)}}{{\mathrm{sinc}(\delta) - N\xi_0 \theta^\delta}}$ and straight line $q = 1$. The abscissa of the red dot is indicated by $B_0$.

 \begin{figure}[!ht]
\centering
\includegraphics[width=0.75\columnwidth]{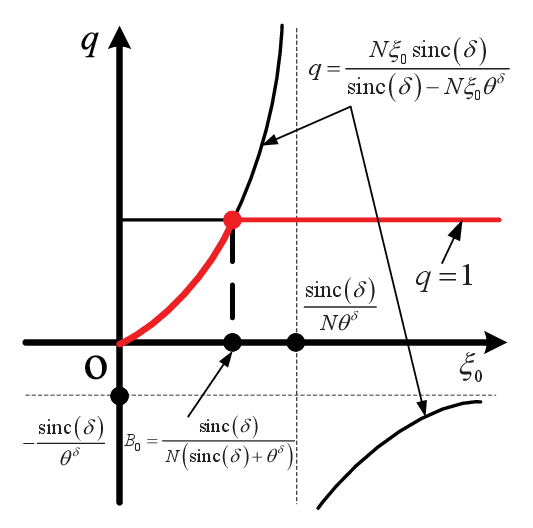} 
\caption{Busy probability in steady state as functions of packet arrival rate ${\xi}_0$. The red curve plots the solution for the fixed-point equation given by \eqref{equation}.}
\label{fig:xy_label1}
\end{figure}

\begin{figure}
  \centering
  \subfigure[The case where $0 < {\xi _0} < \frac{{\mathrm{sinc}(\delta)}}{{N\left( {\mathrm{sinc}(\delta)+ {\theta^\delta}} \right)}}$.]{
    \label{fig:xy_label2} 
    \includegraphics[width=0.8\columnwidth]{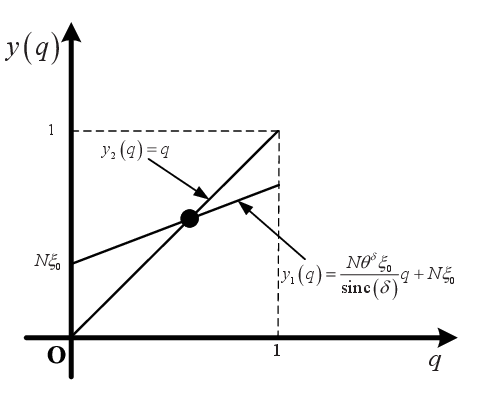}}
  \subfigure[The case where ${\xi _0} \ge \frac{\mathrm{sinc}(\delta)}{N(\mathrm{sinc}(\delta)+ \theta^\delta)}$.]{
    \label{fig:xy_label3} 
    \includegraphics[width=0.8\columnwidth]{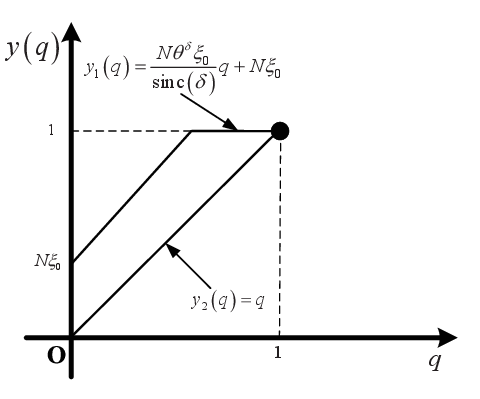}}
  \caption{Illustration of both right-side and left-side for the fixed-point equation given by \eqref{equation} in two cases.}
  \label{fig:xy_label23} 
\end{figure}

\begin{itemize}
\item{Case 1: $0 < \frac{{N\xi_0( {\mathrm{sinc}(\delta)+q\theta^\delta})}}{{\mathrm{sinc} (\delta)}} < 1$.} In this case, we have $0 < q < 1$ and $\frac{N\xi_0( {\mathrm{sinc}(\delta)+ q\theta^\delta})}{\mathrm{sinc}(\delta)} = q$. The solution of \eqref{equation} depends on the intersection of the curve $y_1(q)$ and the curve $y_2(q)$ as shown in Figure \ref{fig:xy_label23}. Therefore, if $0 < {\xi _0} < \frac{\mathrm{sinc} (\delta)}{ N( {\mathrm{sinc}(\delta)+ \theta^{\delta}})}$, the solution of the fixed-point equation is $q = \frac{N{\xi _0}\mathrm{sinc}(\delta) }{\mathrm{sinc}(\delta)- N\xi_0{\theta^\delta}}$; otherwise, the solution becomes $q=1$ if ${\xi_0} \ge \frac{{\mathrm{sinc}(\delta)}}{{N\left( {\mathrm{sinc}(\delta)+ \theta^{\delta}} \right)}}$.

\item{Case 2: $\frac{N\xi_0( {\mathrm{sinc}(\delta) + q\theta^\delta})}{{\mathrm{sinc}(\delta)}} \ge 1$.} In this case, the solution is $q=1$. Then, we get the conclusion that if $\xi_0\ge \frac{{\mathrm{sinc}(\delta)}}{ N( \mathrm{sinc}(\delta)+ \theta^{\delta} )}$, the solution of the fixed-point equation is $q=1$.
\end{itemize}
Thus, we obtain the solution of the fixed-point equation \eqref{approxeq31} in the lemma.
\end{IEEEproof}

\subsection{Success Probability}
Having derived the busy probability $q$, plugging \eqref{equation32} into \eqref{equation34}, we get the approximated success probability as
\begin{align}
\mathbf{P}_s=\left\{ {\begin{aligned}
&1 - \frac{{ N{\xi _0}{\theta ^{\delta}}}}{{\mathrm{sinc}(\delta)}},& &{\rm{if} \  0 < {\xi _0} < \frac{\mathrm{sinc} (\delta)}{\textit{N}(\mathrm{sinc}(\delta) + \theta^\delta)}}\\
&\frac{{\mathrm{sinc} (\delta)}}{{\mathrm{sinc} (\delta) + {\theta ^{\delta}}}},& &{\rm{if} \  {\xi _0} \ge \frac{{\mathrm{sinc} (\delta)}}{{\textit{N}\left( {\mathrm{sinc} (\delta)+ {\theta ^{\delta}}} \right)}}}
\end{aligned}} \right. .
\label{equation35}
\end{align}

\begin{rem}
When the packet arrival rate satisfies $0 < {\xi _0} < \frac{\mathrm{sinc}(\delta)}{\textit{N}(\mathrm{sinc}(\delta) + \theta^\delta)}$, the success probability is determined by the path loss exponent, the number of users and the SIR threshold. When $\xi_0$ or $N$ increases, the success probability decreases due to the increment of traffic. When the path loss exponent increases, the success probability increases, which is attributed to the reduction of the interference. As the SIR threshold increases, the success probability decreases gradually because the SIR requirement becomes higher but the propagation environment is unchanged. On the other hand, when the packet arrival rate satisfies ${\xi _0} \ge \frac{\mathrm{sinc}(\delta)}{\textit{N}(\mathrm{sinc}(\delta) + \theta^\delta)}$, the success probability is only relies on the path loss exponent and the SIR threshold. As the path loss exponent increases, the success probability increases gradually and reaches a maximum value, which implies that the interference from other BSs is dominant for large arrival rate, and the effect of path loss on the interference is more evident than that on the signal.
\end{rem}

\subsection{Throughput}
In order to take both success probability and frequency of transmission into consideration, we derive the throughput per unit spectrum (bps/Hz) of a BS in the following discussions. In the description of system model, we have assumed that the size of packets is fixed to be the same, and a BS requires exactly one time slot to deliver one packet. Since a packet is transmitted by a BS in any time slot with probability $q$ and successful delivered with probability $\mathbf{P}_s$, we get the throughput per unit spectrum of a BS as
\begin{align}
\tau =q\mathbf{P}_s \frac{{L_p }}{{B \cdot \Delta t}},
\end{align}
where $L_p$ is the length of one packet, $B$ is the bandwidth and $\Delta t$ is the length of one time slot.
Assume that as long as the SIR exceeds the threshold $\theta$, a link could transmit at the rate $\log_2(1+\theta)$ bits per second per Hz, i.e., $\frac{L_p}{B\cdot\Delta t} = \log_2(1+\theta)$.

Using \eqref{equation32} and \eqref{equation35}, we get the throughput at each BS as
\begin{align}
\tau  = \left\{ {\begin{array}{ll}
{N\xi_0\log_2(1+\theta)},&{\rm{if} \ 0 < {\xi _0} < \frac{\mathrm{sinc}(\delta)}{N(\mathrm{sinc}(\delta)+ \theta^\delta)}}\\
\frac{\mathrm{sinc}(\delta) \log_2(1+\theta)}{\mathrm{sinc}(\delta) + \theta^\delta},&{\rm{if} \ {\xi_0} \ge \frac{\mathrm{sinc}(\delta) }{N(\mathrm{sinc} (\delta) + \theta ^\delta)}}
\end{array}} \right..
\label{tao}
\end{align}

Observing the equation \eqref{tao}, we can find that the throughput of a BS is related to the path loss exponent, the distribution of users and the threshold of SIR. The throughput increases linearly with $N$ or $\xi_0$, for $0 < {\xi _0} < \frac{\mathrm{sinc}(\delta)}{N(\mathrm{sinc}(\delta)+ \theta^\delta)}$. The linear increase comes from the increase of users service requests. When the packet arrival rate satisfies the condition ${\xi _0} \ge \frac{\mathrm{sinc}(\delta)}{N(\mathrm{sinc}(\delta)+ \theta^\delta)}$, the throughput will only related to the path loss exponent and the threshold of SIR, which means the BS reaches its maximum capacity and the BS will be active all the time.

\begin{figure}[!ht]
\centering
\includegraphics[width=1\columnwidth]{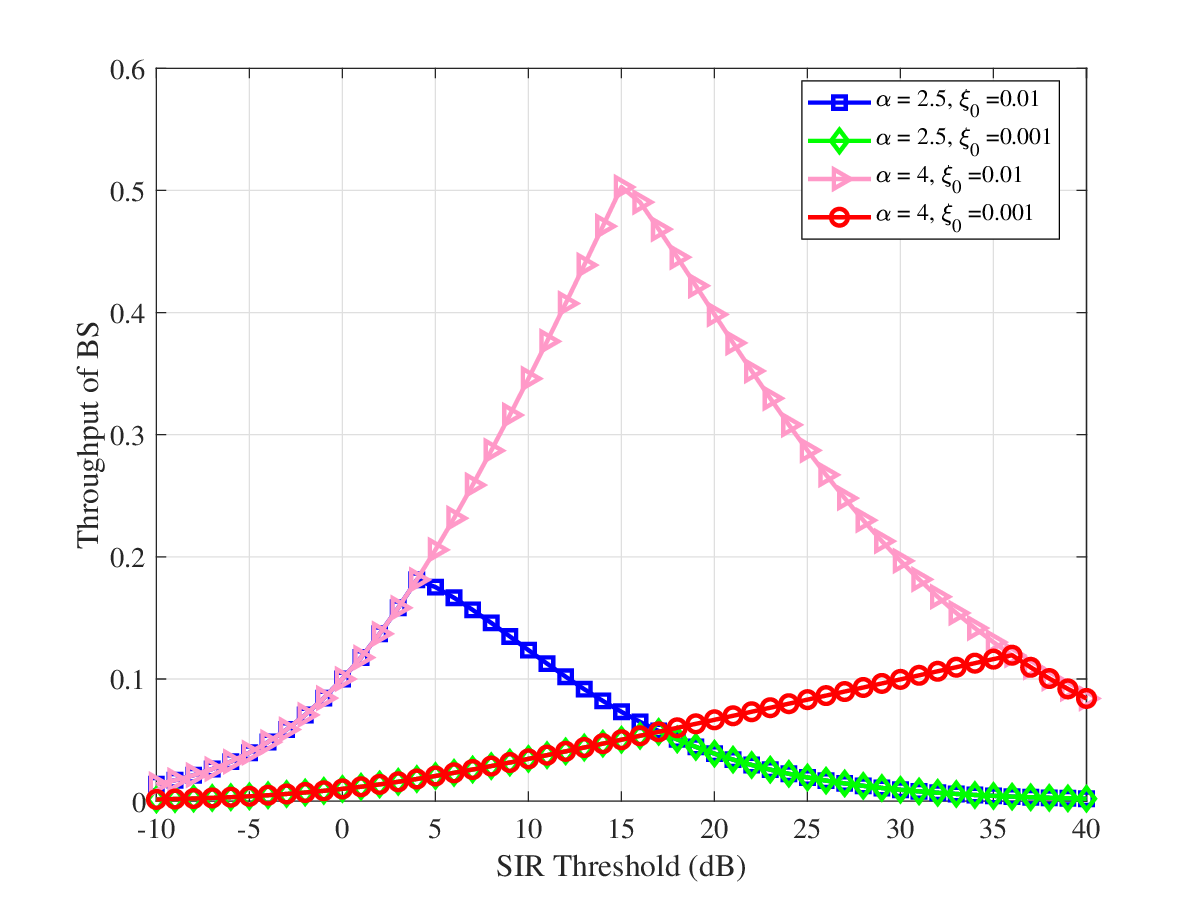} 
\caption{Throughput for different path loss exponents and packet arrival rates. The number of users is $N=10$.}
\label{fig:tao_theta}
\end{figure}

In Figure \ref{fig:tao_theta}, we plot the throughput $\tau$ with respect to the SIR threshold $\theta$ (dB) for different path loss exponents and packet arrival rates . As it can be observed from the curves, it seems counterintuitive that the throughput increases when the path loss exponent increases. The increase comes from the increase of success probability, and it means that the large path loss exponent is not always bad for the throughput. When $0 < {\xi _0} < \frac{\mathrm{sinc}(\delta)}{N(\mathrm{sinc}(\delta)+ \theta^\delta)}$ holds and the packet arrival rate is the same, the throughput for different path loss exponents will be the same. In this case, the packet arrival rate will be the dominant factor that affects the throughput. When ${\xi _0} \ge \frac{\mathrm{sinc}(\delta)}{N(\mathrm{sinc}(\delta)+ \theta^\delta)}$ and the pass loss exponents is the same, the throughput for different packet arrival rates will be the same as well. Similarly, the path loss exponents are the dominant factor to affect the throughput in this case. Observing the curves, we observe that the peak values of the throughput are determined by the path loss exponent and the packet arrival rate. When the path loss exponent and the packet arrival rate increase, the peak value of the throughput will increase. For different path loss exponents and packet arrival rates, the optimal SIR thresholds that achieves the maximum throughput are different.

\subsection{Mean Delay}
In the following, we discuss the mean delay for random scheduling of active users and analyze the relationship between the delay performance and the traffic. Similar to the previous derivations of busy probability, we also consider a single large queue at each BS. For random scheduling of active users, the mean delay of each user equals the mean delay of packets in the large queue at BSs. For such queueing system, the packet arrival rate is $N\xi_0$, and the service rate $\mu$ equals the success probability $\mathbf{P}_s$, which is
\begin{align}
\mu= \left\{ {\begin{aligned}
&1-\frac{N\xi_0\theta^\delta}{\mathrm{sinc}(\delta)},& &{\rm{if} \ 0 < {\xi _0} < \frac{\mathrm{sinc}(\delta)}{N(\mathrm{sinc}(\delta) + \theta^\delta)}}\\
&\frac{\mathrm{sinc}(\delta)}{\mathrm{sinc}(\delta) + \theta^\delta},& &{\rm{if} \ {\xi _0} \ge \frac{{\mathrm{sinc}(\delta) }}{N({\mathrm{sinc}(\delta) + \theta^\delta})}}
\end{aligned}} \right. .
\label{equation37}
\end{align}

Due to the independence of transmissions in different time slots, the probability for successfully transmitting a packet in any time slot is $\mu$, and the service time of each packet is an random variable with geometric distribution. Since the arrival process of packets intended for each user is a Bernoulli process, we could use a discrete-time Geo/Geo/1 queueing model \cite{Atencia2004} to describe the queueing system. The following theorem gives the mean delay for random scheduling.

\begin{theorem}
For random scheduling of active users, the mean delay $\mathbf{D _{\xi _0}}$ is
\begin{align}
\mathbf{D _{\xi _0}} = \left\{ \begin{array}{ll}
\frac{(1 - N\xi_0)\mathrm{sinc}(\delta)}{\mathrm{sinc}(\delta) - N\xi_0\theta^\delta - N{\xi_0}\mathrm{sinc}(\delta) },&\rm{if} \  N < \frac{\mathrm{sinc}(\delta)}{\xi_0(\mathrm{sinc}(\delta) + \theta^\delta)}\\
\infty,&\rm{if} \  N\ge\frac{\mathrm{sinc}(\delta)}{\xi_0(\mathrm{sinc}(\delta) + \theta^\delta)}
\end{array} \right. .
\label{equation38}
\end{align}

\end{theorem}
\begin{IEEEproof}
From \cite[Corollary 2 eq.7] {Atencia2004}, the mean time that a customer spends in the system is
\begin{align}
W = {\beta _1} + \frac{{2\bar p\left( {{\beta _1} - 1} \right)\left[ {1 - A\left( {\bar p} \right)} \right] + p{\beta _2}}}{{2\left[ {p + \bar pA\left( {\bar p} \right) - \rho } \right]}},
\end{align}
where $\beta_n$ is the $n$th factorial moment of service rate, $p$ is the arrival rate ($\bar p = 1 - p$), $\rho = p\beta_1$ is the traffic intensity, $A\left( x \right) = \sum\nolimits_{i = 0}^\infty  {a_i x^i }$ is the Ordinary generating function of successive interretrial times of any user with an arbitrary distribution $\left\{ {a_i } \right\}_{i = 0}^\infty$.
The retrial time is $0$ in our queue system, so we have $a_0  = 1, a_{i \ge 1}  = 0$ and ${A\left( {\bar p} \right)}=1$. The service rate follows from geometric distribution.
Combined with our own queue system, we get $\beta _1 = 1/\mu$, $\beta _2 = 2/{\mu ^2} - 2/\mu$, $p = \xi _0$, ${A\left( {\bar p} \right)}=1$ and $\rho = p \beta _1$. By utilizing the above formulas, plugging in the arrival rate $N\xi_0$ and the service rate $\mu$, we get the mean delay as
\begin{align}
\mathbf{D_{\xi_0}} = \left\{\begin{aligned}
&\frac{1-N\xi_0}{\mu-N\xi_0},& &\rm{if} \ \mu  > N\xi_0\\
&\infty,& &\rm{if} \ \mu  \le N\xi_0
\end{aligned}\right. .
\label{equation40}
\end{align}
Plugging \eqref{equation37} into \eqref{equation40}, we get the result in \eqref{equation38}.
\end{IEEEproof}

\begin{rem}
In the previous discussions of busy probability, success probability and throughput, we find that these performance metrics for random scheduling of active users is the same as that for random scheduling of all users. However, the two different scheduling schemes are distinct in terms of mean delay. For random scheduling of all users, each user is scheduled for transmission in a time slot with probability $1/N$, and the service rate of the typical user is $\mu=\mathbf{P}_s/N$. Then, we get the mean delay as
\begin{align}
\mathbf{D_{\xi_0}} &= \left\{\begin{aligned}
&\frac{1-\xi_0}{\mu-\xi_0},& &\rm{if} \ \mu  > N\xi_0\\
&\infty,& &\rm{if} \ \mu  \le N\xi_0
\end{aligned}\right. \nonumber \\
&=\left\{ \begin{array}{ll}
\frac{(1 - \xi _0) N\mathrm{sinc}(\delta)}{\mathrm{sinc}(\delta) - N\xi_0\theta^\delta - N{\xi_0}\mathrm{sinc}(\delta) },&\rm{if} \  N < \frac{\mathrm{sinc}(\delta)}{\xi_0(\mathrm{sinc}(\delta) + \theta^\delta)}\\
\infty,&\rm{if} \  N\ge\frac{\mathrm{sinc}(\delta)}{\xi_0(\mathrm{sinc}(\delta) + \theta^\delta)}
\end{array} \right. .
\end{align}
Comparing the above result with that in Theorem 1, we observe that the mean delay for random scheduling of all users is much larger than that for random scheduling of active users.
\end{rem}

When the mean delay is finite and $\mathbf{D _{\xi _0}} < \beta$, we get the following inequality
\begin{align}
\frac{{\left( {1 - N{\xi _0}} \right) \mathrm{sinc}\left( {\delta} \right)}}{{\mathrm{sinc}\left( {\delta} \right) -  N\xi_0\theta^\delta - N{\xi_0}\mathrm{sinc}(\delta)}} < \beta.
\end{align}
After simplifying the above inequality, we obtain
\begin{align}
N < \frac{(\beta-1)\mathrm{sinc}(\delta)}{\xi_0(\theta^\delta\beta + \beta\mathrm{sinc}(\delta)-\mathrm{sinc}(\delta))}.
\end{align}
In order to facilitate the above expression, we define
\begin{align}
{A_1} \buildrel \Delta \over = \frac{\mathrm{sinc}(\delta)}{\xi_0({\theta ^{\delta}} + \mathrm{sinc}(\delta))},
\end{align} 
\begin{align}
{A_2} \buildrel \Delta \over = \frac{(\beta-1)\mathrm{sinc}(\delta)}{\xi_0(\theta^\delta\beta + \beta\mathrm{sinc}(\delta)-\mathrm{sinc}(\delta))},
\end{align}
where ${A_1} > {A_2}$ and $\beta$ is the delay requirement for users.

By analyzing the above inequalities, we obtain several conclusions as follows.

\begin{itemize}
  \item When the number of users $N$ satisfies $N < A _2$, the queue is stable, and the delay requirement can be satisfied. By observing the expression of $A _2$, we obtain that the larger the value of $\beta$ is, the greater the value that $N$ will be. In other words, when the delay requirements are low, the typical cell can accommodate more users and satisfies their delay requirements.
  \item When $N$ satisfies the condition $A _2 < N < A _1$, the queue is stable but the delay requirements of users cannot be satisfied, i.e. users can be served by the associated BS, but their delay requirements cannot be met.
  \item When $N$ satisfies the condition $N > A _1$, the queue is not stable, i.e., the users will be blocked in the system and can not be successfully served by the associated BS.
\end{itemize}

\subsection{Stability of Queues}
In this subsection, we derive the unstable probability of the queue for users, which reveals the stability of queues in the wireless networks. A system is said to be stable if its long run averages exist and finite. If a system is unstable, its long run measures are meaningless.

\begin{definition}
For a queueing system with packet arrival rate $\xi$ and service rate $\mu$, the queue is stable if and only if the arrival rate $\xi$ is less than the service rate $\mu$. For a network with multiple queues, the proportion of stable queues among all queues equals to the stable probability of the queue at the typical link, which is defined as
\begin{align}
\mathbf{P}_{\rm{stable}} \buildrel \Delta \over = \mathbb{P}\{\xi < \mu \}.
\end{align}
\end{definition}

With the definition of stable probability, the corresponding unstable probability is given by $\mathbf{P}_{\rm{us}} = \mathbb{P}\{ \xi \ge \mu \}$. Therefore, we get the following theorem.
\begin{theorem}
The unstable probability of queues for users in the network is
\begin{align}
{\mathbf{P}_{\rm{us}}} = 1 - \sum\limits_{k = 1}^\infty  {\mathbb{P}\{ \xi_i \le f(k) \}} \mathbb{P}\{ {N = k} \} - \mathbb{P}\{ {N = 0} \},
\label{equation45}
\end{align}
where $f(k) = \frac{\mathrm{sinc}(\delta)}{k({\mathrm{sinc}(\delta) + \theta^\delta})}$, and $\mathbb{P}\{ N = k \}$ is the PMF of $N$ given by \eqref{equation13} and \eqref{equation14}.
\end{theorem}

\begin{IEEEproof}
The unstable probability can be derived as
\begin{align}
\mathbf{P}_{\rm{us}} &= \mathbb{P}\Big\{ N \ge \frac{\mathrm{sinc}(\delta)}{\xi_i(\mathrm{sinc}(\delta) + \theta^\delta)} \Big\}
\nonumber \\
&= 1 - \sum\limits_{k = 1}^\infty {\mathbb{P}\{\xi_i \le f(k) \}} \mathbb{P}\{ {N = k} \} - \mathbb{P}\{N=0\},
\end{align}
where $\mathbb{P}\{ {{\xi_i} \le f(k)} \}$ can be evaluated by the cumulative distribution function of $\xi_i$, and $\mathbb{P}\{N = k\}$ can be obtained by \eqref{equation13} or \eqref{equation14}.
\end{IEEEproof}

First, we consider the exponentially distributed case where ${\xi _i} \sim \rm{Exp}(\lambda)$,
the unstable probability can be derived as
\begin{align}
{\mathbf{P}_{\rm{us}}} &= 1 - \sum\limits_{k = 1}^\infty  {\Big( 1 - \exp\Big( - \frac{\lambda \mathrm{sinc}(\delta)}{ k( \mathrm{sinc} (\delta) + \theta^\delta )}\Big) \Big)} \mathbb{P}\{N=k\} \nonumber \\
&- \mathbb{P}\{N = 0\}.
\label{Pus_E}
\end{align}

For the uniformly distributed case where ${\xi _i} \sim \rm{Uni}(0,b)$, the unstable probability is
\begin{align}
{\mathbf{P}_{\rm{us}}} &= 1 - \sum\limits_{k = a}^\infty {\frac{{\mathrm{sinc}(\delta)}\mathbb{P}\{N=k\}}{{ kb( {\mathrm{sinc}(\delta) + {\theta ^{\delta}}} )}}} - \sum\limits_{k = 0}^{a - 1} {\mathbb{P}\{N = k\}},
\label{Pus_U}
\end{align}
where the condition is $\frac{{{\rm{sinc}}( {\delta} )}}{{ a( \rm{sinc}(\delta) + \theta ^\delta  )}} \le b < \frac{{{\rm{sinc}}(\delta)}}{{(a - 1)( {{\rm{sinc}}(\delta) + \theta ^{\delta} } )}}$.

When $\xi_i\sim \rm{Exp}(\lambda)$ and the users form a PPP, combined \eqref{equation13} with \eqref{Pus_E}, we get
\begin{align}
\mathbf{P}_{\rm{us}}  = 1 - \sum\limits_{k = 1}^\infty {\Big( {1 - e^{ - \frac{{\lambda\mathrm{sinc} (\delta)}}{{ k( {\mathrm{sinc}(\delta) + \theta ^{\delta} } )}}} } \Big)} \frac{{e^{ - \lambda _u S} }}{{k!}}( {\lambda _u S})^k  - e^{ - \lambda _u S}.
\label{Pus_E_PPP}
\end{align}

When the users are distributed as a PCP, combined \eqref{equation14} with \eqref{Pus_E}, the unstable probability is
\begin{align}
\mathbf{P}_{\rm{us}}  &= 1 - \sum\limits_{k = 1}^\infty  \Bigg( {\bigg( {1 - e^{ - \frac{{\lambda \mathrm{sinc} \left( {\delta} \right)}}{{ k\left( {\mathrm{sinc} \left( {\delta} \right) + \theta ^{\delta} } \right)}}} } \bigg)} \Bigg.\nonumber \\
&\Bigg.\times \bigg( {\sum\limits_{a = 0}^\infty  {\frac{{e^{ - \lambda _p S} }}{{a!}}\left( {\lambda _p Se^{ - \lambda _c \pi r_c^2 } } \right)^a \frac{{\left( {\lambda _c a\pi r_c^2 } \right)^k }}{{k!}}} } \bigg)\Bigg)  \nonumber \\
&- \exp\Big(( {e^{ - \lambda _c \pi r_c^2 }  - 1} )\lambda _p S\Big).
\label{Pus_E_PCP}
\end{align}

When ${\xi _i} \sim \rm{Uni}\left( {0,b} \right)$ and the users form a PPP, combined \eqref{equation13} with \eqref{Pus_U}, we get
\begin{align}
\mathbf{P}_{\rm{us}}  &= 1 - \sum\limits_{k = m}^\infty  {\left( {\frac{{\mathrm{sinc} (\delta)}}{{ kb( {\mathrm{sinc}(\delta) + \theta^\delta })}}} \right)\frac{{e^{ - \lambda _u S} }}{k!}( \lambda _u S)^k } \nonumber \\
&- \sum\limits_{k = 0}^{m - 1} {\frac{{e^{ - \lambda_u S} }}{{k!}}\left( {\lambda _u S} \right)^k }.
\label{Pus_U_PPP}
\end{align}

Similarly, when the users are distributed as a PCP, combined \eqref{equation14} with \eqref{Pus_U}, we get
\begin{align}
  \mathbf{P}_{\rm{us}}  &= 1 - \sum\limits_{k = m}^\infty   \bigg({\frac{{\mathrm{sinc}(\delta)}}{kb( {\mathrm{sinc}(\delta) + \theta^\delta } )}} \Bigg.\nonumber \\
  &\Bigg.\times  {\sum\limits_{a = 0}^\infty  {\frac{{e^{ - \lambda _p S} }}{{a!}}\left( {\lambda _p Se^{ - \lambda _c \pi r_c^2 } } \right)^a \frac{{\left( {\lambda _c a\pi r_c^2 } \right)^k }}{{k!}}} } \bigg)
 \nonumber \\
 & - \sum\limits_{k = 0}^{m - 1} {\left( {\sum\limits_{a = 0}^\infty  {\frac{{e^{ - \lambda _p S} }}{{a!}}( {\lambda_p Se^{ - \lambda _c \pi r_c^2 } } )^a \frac{{(\lambda_c a\pi r_c^2)^k}}{{k!}}} } \right)}.
 \label{Pus_U_PCP}
\end{align}
The numerical results of unstable probability are shown in Figure \ref{fig:Pus_E_natu_natc}, Figure \ref{fig:Pus_U_natu_natc} and Figure \ref{fig:Pus_EUCC_natu}.

\section{Numerical Evaluation}
\label{sec6}
\subsection{Delay Performance}
\begin{figure}[!ht]
\centering
\includegraphics[width=1\columnwidth]{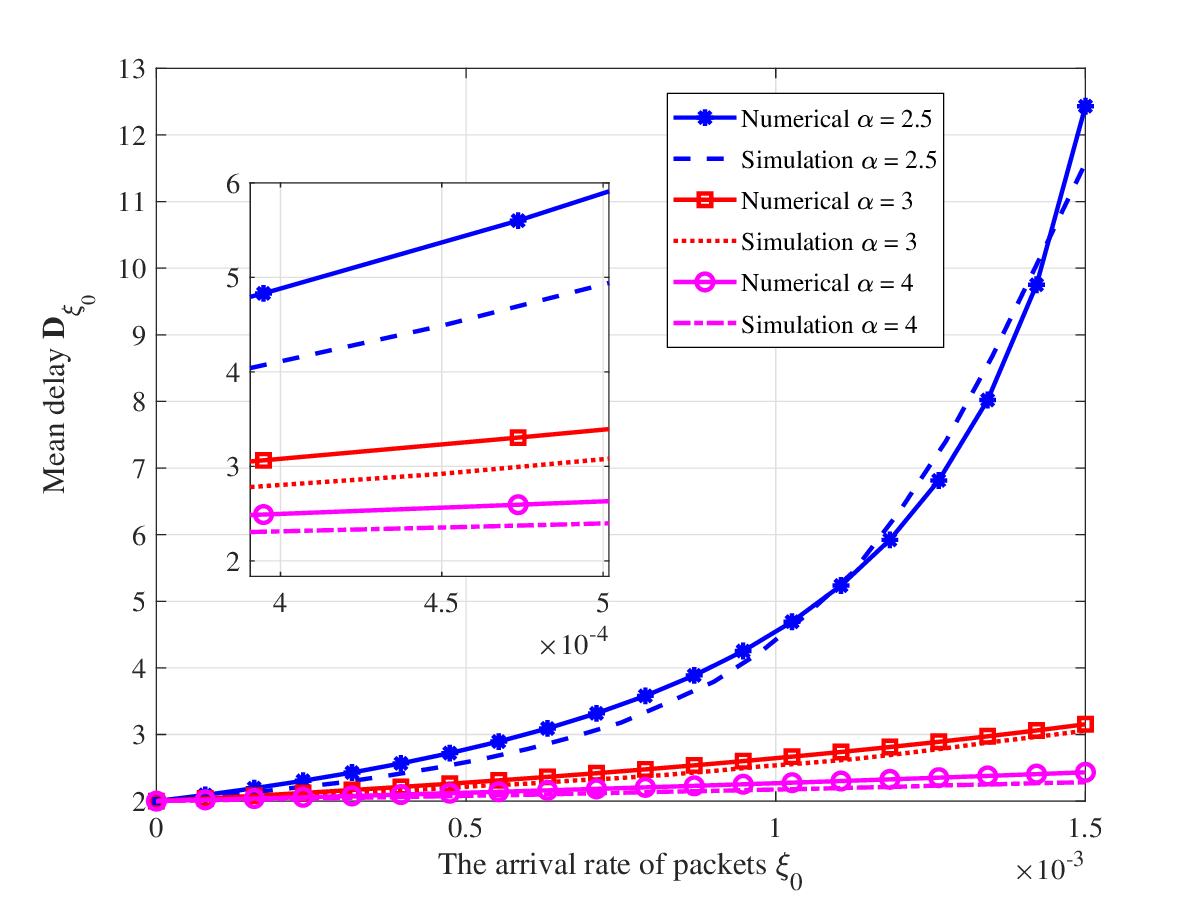} 
\caption{{Effect of the arrival rate $\xi_0$ on the conditional mean delay $\mathbf{D}_{\xi_0}$ for different $\alpha$ ($N = 20, \theta = 10$).}}
\label{fig:Dkesi_kesi}
\end{figure}

In Figure \ref{fig:Dkesi_kesi}, we plot the conditional mean delay $\mathbf{D _{\xi _0}}$ given by \eqref{equation38} as functions of the packet arrival rate $\xi_0$ for different path loss exponents. We observe that, when the value of $\xi_0$ is small, as $\xi_0$ increases, the conditional mean delay $\mathbf{D _{\xi _0}}$ increases gradually since the waiting time is longer due to the increase of the number of arrival packets in a slot time. When the value of $\xi_0$ is larger than a certain value, the conditional mean delay $\mathbf{D _{\xi _0}}$ will be infinite which means that the queue is unstable. This can be interpreted as that the service ability of BSs are limited and the system will be blocked when the traffic is overloaded in each time slot. As the path loss exponent $\alpha$ increases, the conditional mean delay $\mathbf{D _{\xi _0}}$ decreases since the waiting time is shorter due to the decrease of the number of arrival packets.

\subsection{Unstable Probability}
\begin{figure}[!ht]
\centering
\includegraphics[width=1\columnwidth]{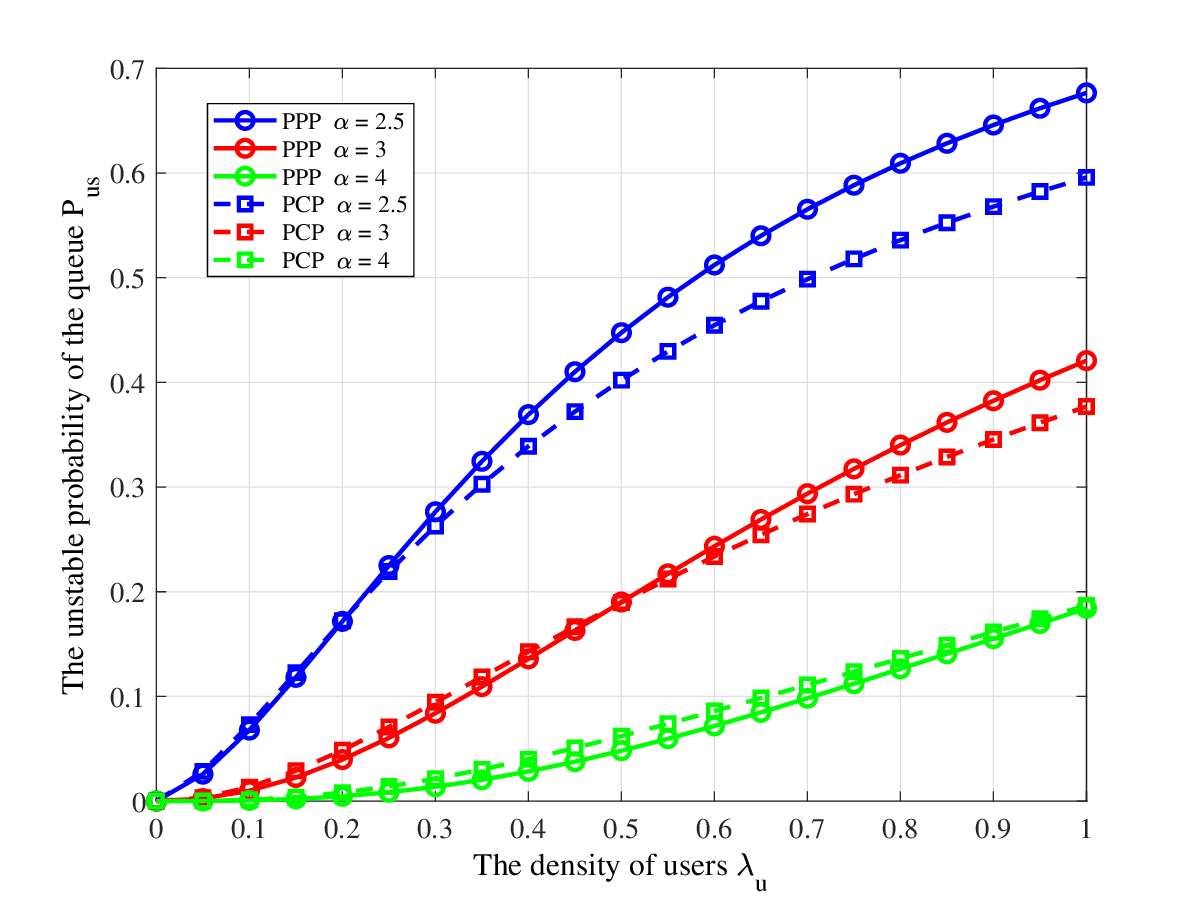} 
\caption{Effect of the density of users $\lambda_u$ on the unstable probability ${\mathbf{P}_{\rm{us}}}$ for different path loss exponents $\alpha$ and various distributions of users when the arrival rate is exponentially distributed as Exp(0.01). The parameters are set as $\theta = 10, S = 10, r_c = 1, \lambda_p = \frac{1}{{1.1\pi }}$ and $\lambda_c = 1.1\lambda_u$.}
\label{fig:Pus_E_natu_natc}
\end{figure}

In Figure \ref{fig:Pus_E_natu_natc}, we plot the unstable probability ${\mathbf{P}_{\rm{us}}}$ given by \eqref{Pus_E} as functions of the density of users $\lambda_u$ for different path loss exponents $\alpha$. The packet arrival rate  $\xi_i$ follows an exponential distribution with mean $0.01$. As the density of users $\lambda_u$ increases, the unstable probability ${\mathbf{P}_{\rm{us}}}$ increases due to the increase of the number of users. When the density of users $\lambda_u$ increases to a large value, the unstable probability ${\mathbf{P}_{\rm{us}}}$ approaches to one.
As shown in Figure 13, when the density of users is small, the unstable probability in non-uniformly distributed case is larger than that in uniformly distributed case. Nevertheless, when the density of users is large, it is reversed that the unstable probability in non-uniformly distributed case become smaller than that in uniformly distributed case. The observation can be interpreted as that when the density of users is small, if users are uniformly distributed in space, the probability of a BS being overburdened is small. However, if users are non-uniformly distributed when the density of users is small, the probability of a BS being overburdened increases. Therefore, in the case of small density of users, the unstable probability of the queues in uniformly distributed case is smaller than that in non-uniformly distributed case. On the other hand, when the density of users is large, if users are uniformly distributed in space, almost most BSs in the network are overburdened, and the proportion of BSs with unstable queues is large. However, if users are non-uniformly distributed in space, the amount of traffic at different BSs become polarized, i.e. either extremely heavy or extremely light. Compared with the uniformly distributed case, the number of BSs with light traffic increases, leading to decrease of the unstable probability.

\begin{figure}[!ht]
\centering
\includegraphics[width=1\columnwidth]{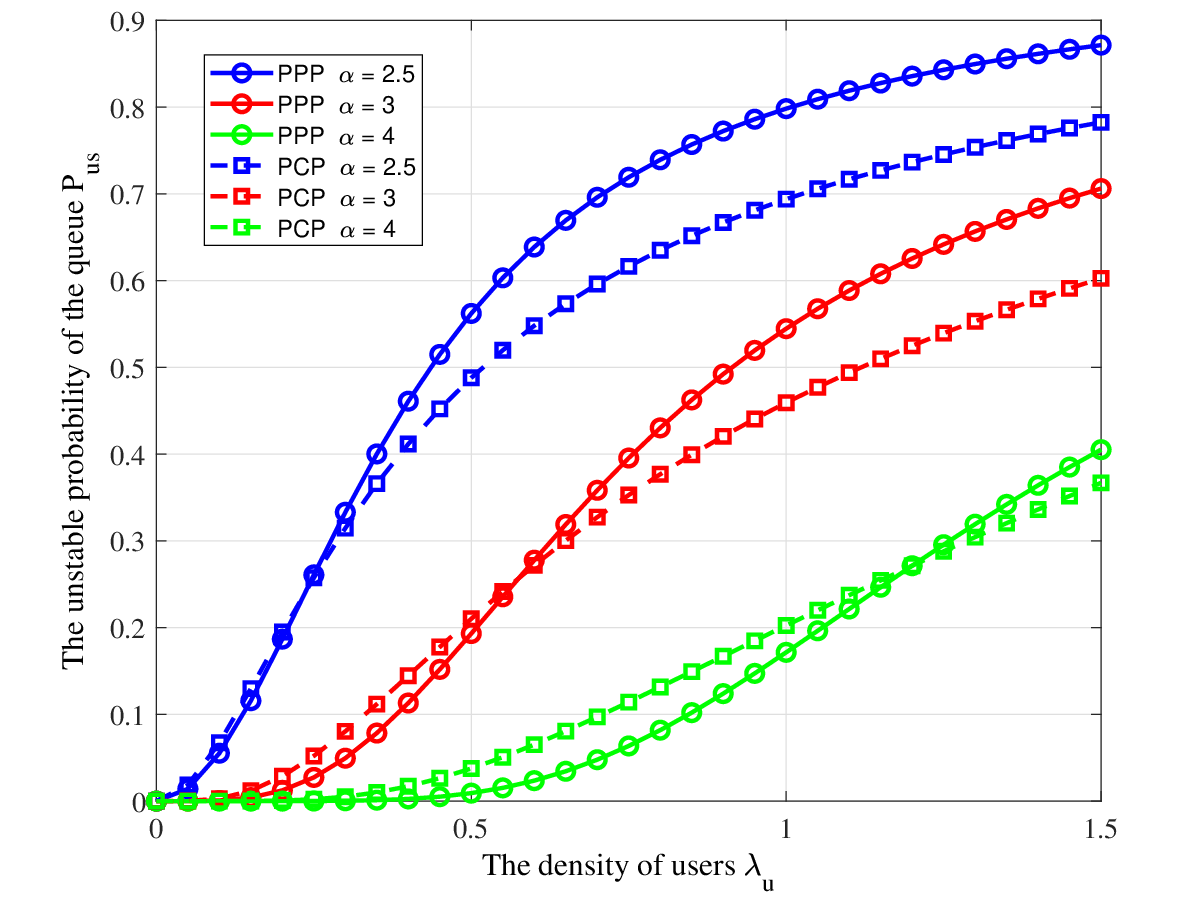} 
\caption{Effect of user density $\lambda_u$ on the unstable probability ${\mathbf{P}_{\rm{us}}}$ for different path loss exponents $\alpha$ and various distributions of users when the arrival rate is uniformly distributed as $\rm{Uni}(0, 0.02)$. The parameters are set as $\theta = 10, S = 10, r_c = 1, \lambda_p = \frac{1}{{1.1\pi }}$ and $\lambda_c = 1.1\lambda_u$.}
\label{fig:Pus_U_natu_natc}
\end{figure}

In Figure \ref{fig:Pus_U_natu_natc}, we plot the unstable probability ${\mathbf{P}_{\rm{us}}}$ given by \eqref{Pus_U} as functions of user density $\lambda_u$ for different path loss exponents $\alpha$. The packet arrival rate $\xi_i$ is uniformly distributed as $\rm{Uni}(0, 0.02)$. Similar to the results in Figure \ref{fig:Pus_E_natu_natc}, as the density of users $\lambda_u$ increases, the unstable probability ${\mathbf{P}_{\rm{us}}}$ increases. As the path loss exponent $\alpha$ increases, the unstable probability ${\mathbf{P}_{\rm{us}}}$ decreases. We can observe a similar trend for the unstable probability as that in Figure \ref{fig:Pus_E_natu_natc}. It further verifies the conclusion that the unstable probability in the uniformly distributed case is smaller than that in the non-uniformly distributed case for small $\lambda_u$, and it is reversed for large $\lambda_u$. Contrast with Figure \ref{fig:Pus_E_natu_natc}, we observe that the unstable probabilities in Figure \ref{fig:Pus_U_natu_natc} are larger than those in Figure \ref{fig:Pus_E_natu_natc} when changing the distribution of $\xi_i$, indicating that the unstable probability is smalller when the arrival rate $\xi_i$ follows an exponential distribution.

\begin{figure}[!ht]
\centering
\includegraphics[width=1\columnwidth]{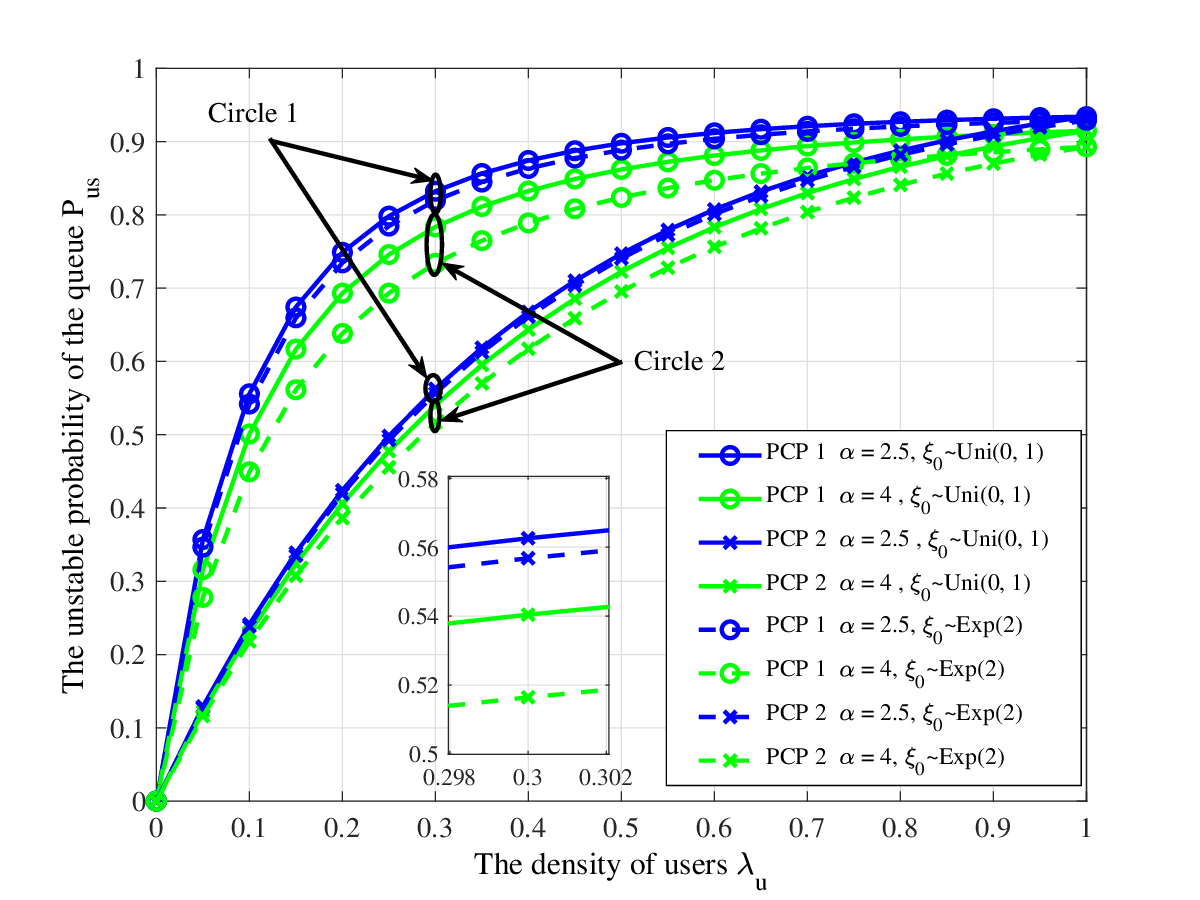} 
\caption{{Effect of user density $\lambda_u$ on ${\mathbf{P}_{\rm{us}}}$ for different path loss exponents $\alpha$ when the arrival rate is either uniformly distributed as $\rm{Uni}(0, 1)$ or exponentially distributed as $\rm{Exp}(0.5)$. Two different settings of parameters are $\theta = 10, S = 10, r_c = 1, \lambda_p = \frac{1}{{1.1\pi}}, \lambda_c = 1.1\lambda_u$ for PCP 1, and $\theta = 10, S = 10, r_c = 1, \lambda_p = \frac{\lambda_u }{{1.1\pi }}, \lambda_c = 1.1$ for PCP 2.}}
\label{fig:Pus_EUCC_natu}
\end{figure}

In Figure \ref{fig:Pus_EUCC_natu}, we plot the unstable probability ${\mathbf{P}_{\rm{us}}}$ given by \eqref{Pus_E} and \eqref{Pus_U} as functions of user density $\lambda_u$ for different distributions of arrival rate.
Two different settings of parameters for PCP, named 'PCP 1' and 'PCP 2', are considered. The packet arrival rate $\xi_i$ is either uniformly distributed as $\rm{Uni}(0, 1)$ or exponentially distributed as $\rm{Exp}(0.5)$. Figure \ref{fig:Pus_EUCC_natu} reveals that the unstable probability in PCP 1 is larger than that in PCP 2.  We also observe that the unstable probability is smaller when $\xi_i$ is exponentially distributed compared to that when  $\xi_i$ is uniformly distributed. In particular, when the path loss exponent $\alpha$ increases from 2.5 to 4, the gaps shown in Circle 1 is enlarged to that in Circle 2, indicating that for medium value of the density of users $\lambda_u$, the unstable probability of queues is not dominated by the probability distribution of $\xi_i$, but by the path loss.

\section{Conclusion}
\label{sec7}
In this paper, we consider a tractable model to analyze the effect of spatio-temporal traffic on the wireless network. By considering a network consisting of one tier of BSs and one tier of users, we compared the distributions of users in the uniformly distributed case and the non-uniformly distributed case and derived the PMF of the number of users, the variance of total arrival rate, the success probability, the throughput, the conditional mean delay and the unstable probability of queue. Specific expressions were obtained for the proposed model. Based on the obtained expressions, we discussed the effect of spatio-temporal traffic on network delay and the unstable probability of queues.

From the numerical evaluation, we observe that the fluctuations of total arrival rate are greater in the non-uniformly distributed case than that in the uniformly distributed case.
The stable probability of queue in the non-uniformly distributed case is larger than that in the uniformly distributed case when the density of users is large or when the arrival rate follows an exponential distribution. Our analyses reveal the difference between the uniformly and non-uniformly distributed traffic and provide insights on the design of wireless networks when various spatio-temporal properties of traffic is considered.

\bibliographystyle{IEEEtran}
\normalem\bibliography{123}

\begin{IEEEbiography}[{\includegraphics[width=1in,height=1.25in,clip,keepaspectratio]{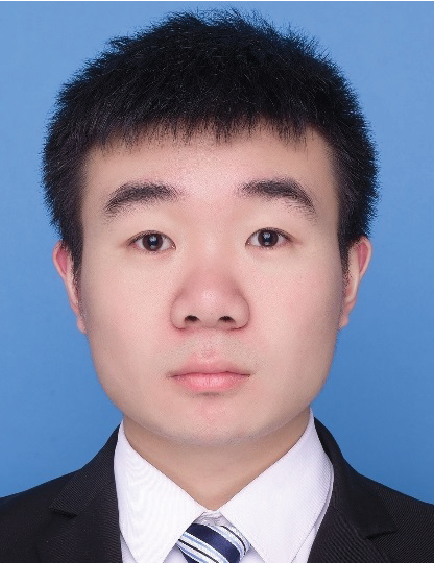}}]{Gang Wang}
(S'18) received the B.S. degree in electronic and information engineering from the Huazhong University of Science and Technology, Wuhan, China, in 2017, where he is currently pursuing the M.S. degree with the School of
Electronic Information and Communication. His research interests include wireless communications, mobile networks, stochastic geometry, and point process theory.
\end{IEEEbiography}

\begin{IEEEbiography}[{\includegraphics[width=1in,height=1.25in,clip,keepaspectratio]{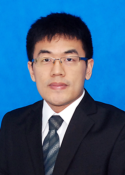}}]{Yi Zhong} (S'12-M'15) received his B.S. and Ph.D. degree in Electronic Engineering from University of Science and Technology of China (USTC) in 2010 and 2015 respectively. From 2015 to 2016, he was a Postdoctoral Research Fellow with the Singapore University of Technology and Design (SUTD) in the Wireless Networks and Decision Systems (WNDS) Group. Now, he is an assistant professor with School of Electronic Information and Communications, Huazhong University of Science and Technology, Wuhan, China. He is an editor of the IEEE Wireless Communications Letters (since 2020), EURASIP on Wireless Communication and Networking (since 2019), Elsevier Physical Communication (since 2019). His main research interests include heterogeneous and femtocell-overlaid cellular networks, wireless ad hoc networks, stochastic geometry and point process theory.
\end{IEEEbiography}

\begin{IEEEbiography}[{\includegraphics[width=1in,height=1.25in,clip,keepaspectratio]{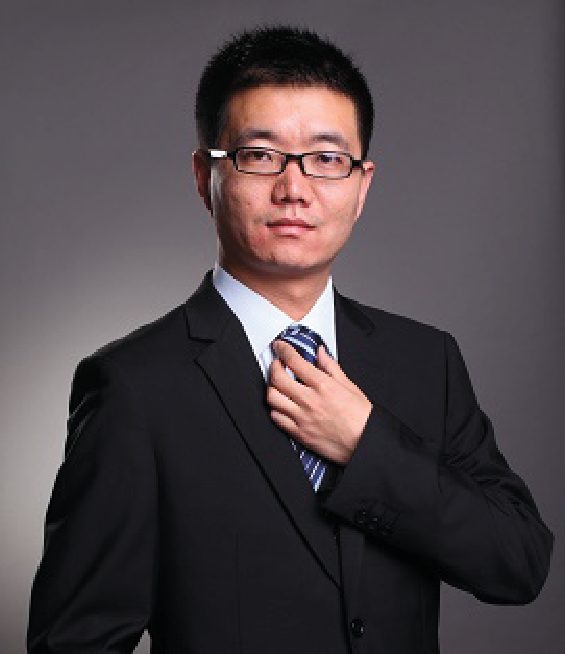}}]{Rongpeng Li}
(S'12-M'15) is now an assistant professor in College of Information Science and Electronic Engineering, Zhejiang University, Hangzhou China. He received his Ph.D and B.E. (with honors) from Zhejiang University, Hangzhou, China and Xidian University, Xi¡¯an, China in June 2015 and June 2010 respectively. From November 2016 to October 2018, Dr. Li was a postdoctoral researcher in College of Computer Science and Technologies, Zhejiang University, Hangzhou, China, which is sponsored by the National Postdoctoral Program for Innovative Talents. From August 2015 to September 2016, he was a research engineer with Wireless Communication Laboratory, Huawei Technologies Co. Ltd., Shanghai, China. He was a visiting scholar in Department of Computer Science and Technology, Cambridge University, UK from Feburary 2020 to August 2020, a visiting doctoral student in Sup¨¦lec, Rennes, France from September 2013 to December 2013, and an intern researcher in China Mobile Research Institue, Beijing, China from May 2014 to August 2014. His research interests currently focus on service-awareness, networking, and artificial intelligence (especially reinforcement learning). He serves as an Editor of China Communications.
\end{IEEEbiography}

\begin{IEEEbiography}[{\includegraphics[width=1in,height=1.25in,clip,keepaspectratio]{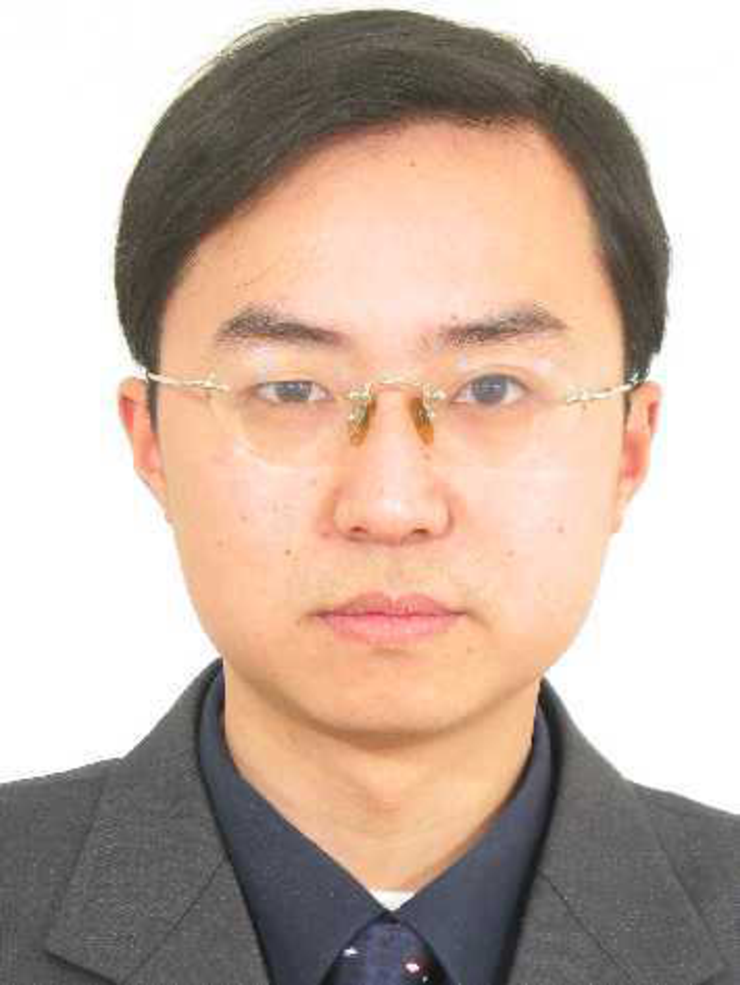}}]{Xiaohu Ge} (M'09-SM'11) received the Ph.D. degree in communication and information engineering from the Huazhong University of Science and Technology (HUST), Wuhan, China, in 2003. He is currently a Full Professor with the School of Electronic Information and Communications, HUST. He is an Adjunct Professor with the Faculty of Engineering and Information Technology, University of Technology Sydney, Ultimo, NSW, Australia. He worked as a Researcher with Ajou University, Suwon, South Korea, and the Politecnico Di Torino, Turin, Italy, from January 2004 to October 2005. He has worked with HUST since November 2005. He has published more than 200 papers in refereed journals and conference proceedings and has been granted about 25 patents in China. His research interests are in the area of mobile communications, traffic modeling in wireless networks, green communications, and interference modeling in wireless communications. Prof. Ge received the Best Paper Awards from IEEE Globecom 2010. He served as the General Chair for the 2015 IEEE International Conference on Green Computing and Communications (IEEE GreenCom 2015). He serves as an Associate Editor for IEEE Wireless Communications, the IEEE Transactions on Vehicular Technology, and IEEE Access.
\end{IEEEbiography}

\begin{IEEEbiography}[{\includegraphics[width=1in,height=1.25in,keepaspectratio]{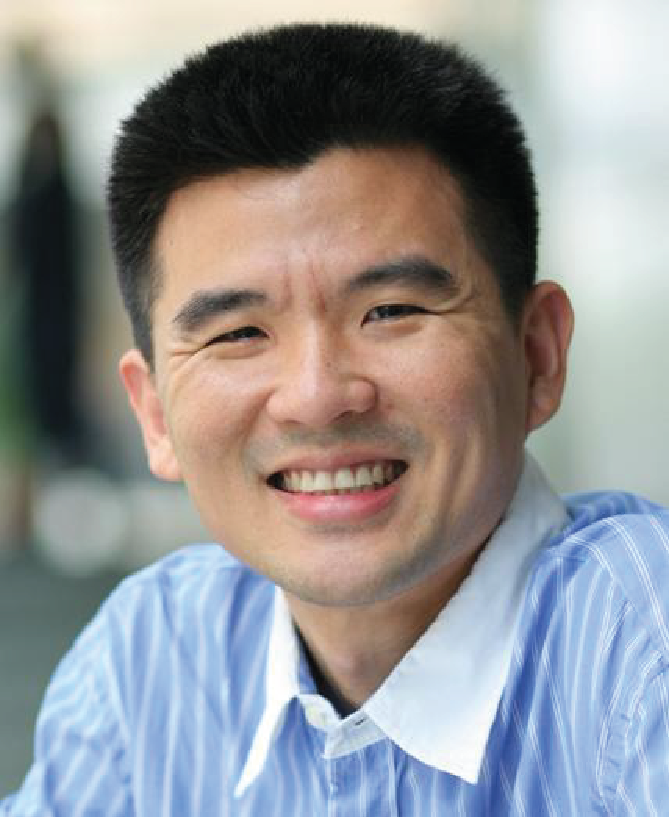}}]
{Tony Q.S. Quek}(S'98-M'08-SM'12-F'18) received the B.E.\ and M.E.\ degrees in electrical and electronics engineering from the Tokyo Institute of Technology in 1998 and 2000, respectively, and the Ph.D.\ degree in electrical engineering and computer science from the Massachusetts Institute of Technology in 2008. Currently, he is the Cheng Tsang Man Chair Professor with Singapore University of Technology and Design (SUTD). He also serves as the Head of ISTD Pillar, Sector Lead of the SUTD AI Program, and the Deputy Director of the SUTD-ZJU IDEA. His current research topics include wireless communications and networking, network intelligence, internet-of-things, URLLC, and big data processing.

Dr.\ Quek has been actively involved in organizing and chairing sessions, and has served as a member of the Technical Program Committee as well as symposium chairs in a number of international conferences. He is currently serving as an Editor for the {\scshape IEEE Transactions on Wireless Communications}, the Chair of IEEE VTS Technical Committee on Deep Learning for Wireless Communications as well as an elected member of the IEEE Signal Processing Society SPCOM Technical Committee. He was an Executive Editorial Committee Member for the {\scshape IEEE Transactions on Wireless Communications}, an Editor for the {\scshape IEEE Transactions on Communications}, and an Editor for the {\scshape IEEE Wireless Communications Letters}.

Dr.\ Quek was honored with the 2008 Philip Yeo Prize for Outstanding Achievement in Research, the 2012 IEEE William R. Bennett Prize, the 2015 SUTD Outstanding Education Awards -- Excellence in Research, the 2016 IEEE Signal Processing Society Young Author Best Paper Award, the 2017 CTTC Early Achievement Award, the 2017 IEEE ComSoc AP Outstanding Paper Award, the 2020 IEEE Communications Society Young Author Best Paper Award, the 2020 IEEE Stephen O. Rice Prize, and the 2016-2019 Clarivate Analytics Highly Cited Researcher. He is a Distinguished Lecturer of the IEEE Communications Society and a Fellow of IEEE.
\end{IEEEbiography}

\begin{IEEEbiography}[{\includegraphics[width=1in,height=1.25in,clip,keepaspectratio]{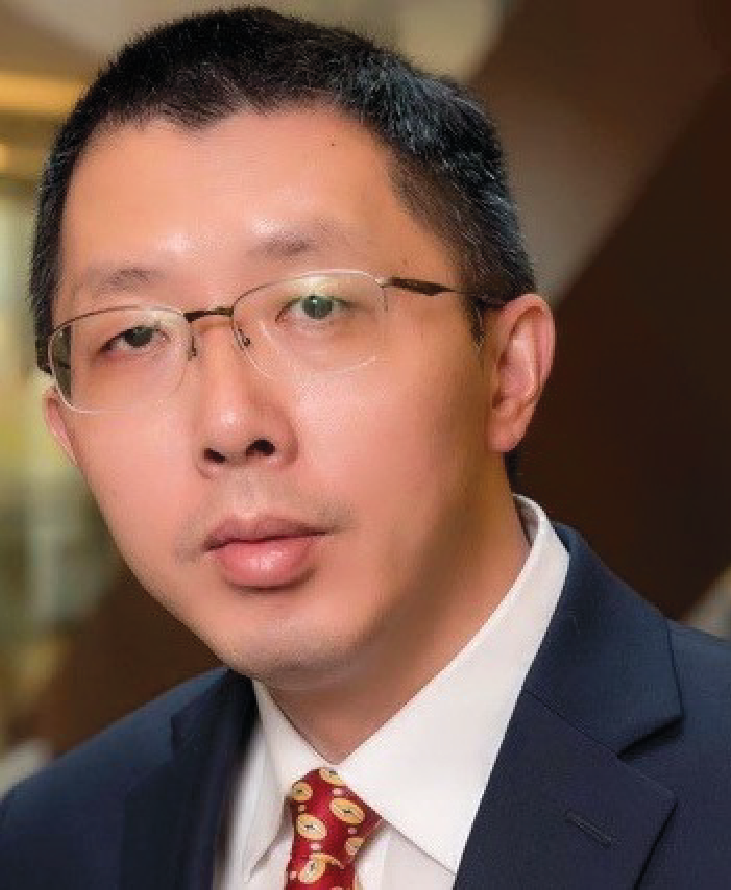}}]{Guoqiang Mao} (S'98-M'02-SM'08-F'18) is a Distinguished Professor at Xidian University. Before that he was with University of Technology Sydney and the University of Sydney. He has published over 200 papers in international conferences and journals, which have been cited more than 9000 times. He is an editor of the IEEE Transactions on Intelligent Transportation Systems (since 2018), IEEE Transactions on Wireless Communications (2014-2019), IEEE Transactions on Vehicular Technology (since 2010) and received ¡°Top Editor¡± award for outstanding contributions to the IEEE Transactions on Vehicular Technology in 2011, 2014 and 2015. He was a co-chair of IEEE Intelligent Transport Systems Society Technical Committee on Communication Networks. He has served as a chair, co-chair and TPC member in a number of international conferences. He is a Fellow of IET. His research interest includes intelligent transport systems, applied graph theory and its applications in telecommunications, Internet of Things, wireless sensor networks, wireless localization techniques and network modeling and performance analysis.
\end{IEEEbiography}

\end{document}